\newif\ifFigsOn\FigsOntrue

\documentclass[sigconf]{acmart}

\usepackage{subcaption}
\usepackage{pgfplots}
\usepackage{placeins}
\usepackage{array}
\usepackage{siunitx}

\usepackage{natbib}
\usepackage{tikz}
\usepackage[super]{nth}

\usetikzlibrary{positioning}
\usetikzlibrary{calc}

\usetikzlibrary{external}

\pgfplotsset{compat=newest}

\def\figscale{0.8}

\AtBeginDocument{%
  \providecommand\BibTeX{{%
    \normalfont B\kern-0.5em{\scshape i\kern-0.25em b}\kern-0.8em\TeX}}}

\setcopyright{none}
\acmYear{2020}

\acmConference[CARRV '20]{CARRV '18: Fourth Workshop on Computer Architecture Research with RISC-V}{May 30, 2020}{Virtual Workshop}
\renewcommand\footnotetextcopyrightpermission[1]{}
\newcommand{\code}[1]{\texttt{#1}}

\makeatletter
\if@ACM@authordraft
  \newcommand{\Comment}[1]{\textbf{\textsl{#1}}}
  \newenvironment{LongComment}[1] 
    {\begingroup\par\noindent\slshape \textbf{\colorbox{yellow}{Begin Comment[#1]}}\par}
    {\par\noindent\textbf{\colorbox{yellow}{End Comment}}\endgroup\par}
  \newcommand{\FIXME}[1]{\textbf{\textsl{\colorbox{yellow}{FIXME:} #1}}}
  \newcommand{\todo}[1]{\textcolor{red}{TODO: \@#1}}
\else 
  \newcommand{\Comment}[1]{\relax}
  
  \newcommand{\FIXME}[1]{\relax}
  \newcommand{\todo}[1]{\relax}
\fi 
\makeatother

\newcommand{\instbit}[1]{\mbox{\scriptsize #1}}
\newcommand{\instbitrange}[2]{~\instbit{#1} \hfill \instbit{#2}~}

\begin{document}\sloppy

  \renewcommand{\sectionautorefname}{Section}
  \renewcommand{\subsectionautorefname}{Section}
  \renewcommand{\subsubsectionautorefname}{Section}
  \renewcommand{\appendixautorefname}{Appendix}
  \renewcommand{\Hfootnoteautorefname}{Footnote}

\title{Prevention of Microarchitectural Covert Channels on an Open-Source 64-bit RISC-V Core} 

\author{Nils Wistoff}
\affiliation{%
  \institution{ETH~Zurich and~RWTH~Aachen and~HENSOLDT~Cyber~GmbH}
  \streetaddress{Gloriastrasse 35}
  \city{Zurich}
  \country{Switzerland}
  \postcode{8092}
}
\email{nwistoff@ethz.ch}

\author{Moritz Schneider}
\affiliation{%
  \institution{ETH Zurich}
  \streetaddress{Universit\"atstrasse 6}
  \city{Zurich}
  \country{Switzerland}}
\email{moritz.schneider@inf.ethz.ch}

\author{Frank K. G\"urkaynak}
\affiliation{%
  \institution{ETH Zurich}
  \streetaddress{Gloriastrasse 35}
  \city{Zurich}
  \country{Switzerland}
  \postcode{8092}
}
\email{kgf@iis.ee.ethz.ch}

\author{Luca Benini}
\affiliation{%
  \institution{ETH Zurich}
  \streetaddress{Gloriastrasse 35}
  \city{Zurich}
  \country{Switzerland}
  \postcode{8092}
}
\email{lbenini@iis.ee.ethz.ch}

\author{Gernot Heiser}
\affiliation{%
  \institution{UNSW Sydney and Data61 CSIRO}
  \city{Sydney}
  \country{Australia}
}
\email{gernot@unsw.edu.au}

\renewcommand{\shortauthors}{Nils Wistoff, Moritz Schneider et al.}

\begin{abstract}
  Covert channels enable information leakage across security boundaries of the operating system. Microarchitectural covert channels exploit changes in execution timing resulting from competing access to limited hardware resources. We use the recent experimental support for time protection, aimed at preventing covert channels, in the seL4 microkernel and evaluate the efficacy of the mechanisms against five known channels on Ariane, an open-source 64-bit application-class RISC-V core. We confirm that without hardware support, these defences are expensive and incomplete. We show that the addition of a single-instruction extension to the \mbox{RISC-V} ISA, that flushes microarchitectural state, can enable the OS to close all five evaluated covert channels with low increase in context switch costs and negligible hardware overhead. We conclude that such a mechanism is essential for security.
\end{abstract}

\ccsdesc[500]{Security and privacy~Hardware security implementation}
\ccsdesc[500]{Security and privacy~Operating systems security}
\ccsdesc[100]{Computer systems organization~Reduced instruction set computing}

\keywords{covert channels, timing channels, computer architecture,
  microarchitecture, operating systems, system security, time protection}


\maketitle


\section{Introduction}
A covert channel is an information flow that uses a mechanism not intended for information transfer~\cite{Lampson_73}, and thereby violates a system's security policy that the OS is meant to enforce. For example, some untrusted code, such as a mail client, may be given access to secrets but should be \emph{confined} to only communicate with the outside world via an encrypted channel. A covert channel can enable the mailer to leak the raw secrets, bypassing encryption.

Covert channels that utilise OS-managed spatial resources (storage channels) can be eliminated completely, as was proved for the seL4 microkernel~\cite{Murray_MBGBSLGK_13}. Harder to control are channels that target physical quantities not directly managed by the OS, such as processor temperature~\cite{masti2015thermal} or power draw~\cite{khatamifard2019powert}. Particularly dangerous are \emph{timing channels}, which exploit information encoded in the timing of events, as they can be exploited remotely.

Of particular importance are microarchitectural timing channels; these exploit competition for limited hardware resources that are hidden by the instruction set architecture (ISA)~\cite{Ge_YCH_18}. For example, the Spectre attack~\cite{Kocher2018spectre} uses speculation to construct a Trojan from ``gadgets'' in innocent code, with the Trojan leaking arbitrary information through a microarchitectural timing channel. Exploitable resources are those holding state that depends on execution history, which includes caches, TLBs, branch predictors, and prefetchers.

\emph{Time protection}, a set of OS mechanisms complementing the established memory protection, aims to prevent timing channels~\cite{Ge2019a}. However, its proponents also demonstrated that on contemporary hardware full time protection is unachievable, as some exploitable  microarchitectural states cannot be reset by software. They consequently argue that the hardware-software contract must be amended to provide the OS with the mechanisms for resetting exploitable microarchitectural state~\cite{Ge2018}.

In this work, we investigate such mechanisms by implementing them in Ariane, an open-source, application-class, in-order, RISC-V RV64 core. We evaluate their efficacy on five known microarchitectural channels, and the overheads imposed by their use. Specifically, we make the following contributions:

\begin{enumerate}
    \item We measure the capacites of five established microarchitectural covert channels on the unmodified Ariane core, and confirm that they are comparable to those found in high-performance Intel and Arm processors.
    \item We confirm previous observations on Intel and Arm cores that software-only approaches are expensive and ineffective. 
    \item We demonstrate the importance of resetting \emph{all} microarchitectural state by showing that after resetting first-order state (valid bits), secondary state (e.g.\ state bits in the cache replacement policies) can still be exploited.
    \item We propose a new RISC-V \code{fence} instruction, whose argument lets the OS control which state is flushed, and demonstrate that it completely eliminates the studied channels.
\end{enumerate}



\section{Background}

\subsection{Threat Model}

We examine covert-channel leakage under a \emph{confinement} scenario~\cite{Lampson_73}: An untrusted program possesses a secret, and the OS  encapsulates the program's execution in a security domain that only allows communication across defined channels to trusted components (e.g., an encryption service). The untrusted program contains a Trojan that is actively trying to leak the secret via a covert channel. A second, unconfined, and also untrusted security domain contains a spy which is trying to read the secret leaked by the Trojan.

Intentional leakage by a Trojan represents the worst case leakage; if we can prevent it, we also preclude any other leakage through the same channel. In particular, this rules out \emph{side channels}, where instead of a Trojan, the leakage originates from an unwitting victim.

We assume that the Trojan and spy time-share a core, meaning cross-core leakage is out of scope. We only consider microarchitectural timing channels. Covert channels that abuse other characteristics, such as power draw, are out of scope.

\subsection{Exploitable Microarchitectural State}



Exploits of data and instruction caches have been known for decades~\cite{Hu_92}. The cache lines used by the Trojan create a footprint that can be sensed by the spy: It observes the latencies of its own memory accesses, which are high where a cache line has been replaced by the Trojan (see \autoref{s:attack} for details).
TLBs are caches for translation data and can be similarly exploited~\cite{Hund_WH_13}.

The branch predictor also contains caches that can be exploited~\cite{aciiccmez2007predicting}: the branch target buffer (BTB), which caches the destination addresses of indirect jumps, and the branch history table (BHT), which predicts whether conditional branches are taken.

Instruction and data prefetchers contain state machines which accumulate history and can be exploited~\cite{Ge2019a}. However, simple processors such as our Ariane core do not feature prefetching; we therefore do not investigate this channel.

\subsection{Exploiting Covert Channels}
\label{s:attack}

Techniques for exploiting covert channels are well established; for our scenario of intentional leakage, the \emph{prime-and-probe} (P\&P) attack~\cite{percival2005cache} is simple and effective.

\begin{figure}
  \centering
  \ifFigsOn
	\scalebox{\figscale}{\begin{tikzpicture}[x=0.24\linewidth,y=1.2\baselineskip]
  \begin{scope}[local bounding box=state1]
    
    \foreach \x in {0,1,...,5}
    {
      \node[inner sep=1pt] (BL\x) at (0,-\x)   {};
      \node[inner sep=1pt] (UR\x) at (1,-\x+1) {};
      \node (MM\x) at ($(BL\x)!0.5!(UR\x)$) {};
      \draw (BL\x) rectangle (UR\x);
    }

    \node[anchor=west] at ($(MM0)-(0.2,0)$)  {\itshape SPY};
    \node[anchor=east] at ($(MM0)-(0.51,0)$) {0};
    \node[anchor=west] at ($(MM1)-(0.2,0)$)  {\itshape SPY};
    \node[anchor=east] at ($(MM1)-(0.51,0)$) {1};
    \node[anchor=west] at ($(MM2)-(0.2,0)$)  {\itshape SPY};
    \node[anchor=east] at ($(MM2)-(0.51,0)$) {2};
    \node[anchor=west] at ($(MM3)-(0.2,0)$)  {\itshape SPY};
    \node[anchor=east] at ($(MM3)-(0.51,0)$) {3};
    \node[anchor=west] at ($(MM4)-(0.2,0)$)  {\ldots};
    \node[anchor=east] at ($(MM4)-(0.51,0)$) {\ldots};
    \node[anchor=west] at ($(MM5)-(0.2,0)$)  {\itshape SPY};
    \node[anchor=east] (nsets) at ($(MM5)-(0.51,0)$) {$N - 1$};


    \draw ($(BL0|-UR0)+(0.25,0)$) -- ($(BL5)+(0.25,0)$);

    \node (state1right) at ($(UR3)+(0.2,0)$) {};
  \end{scope}

  \begin{scope}[shift={(2.2,0)}, local bounding box=state2]
    
    \foreach \x in {0,1,...,5}
    {
      \node[inner sep=1pt] (BL\x) at (0,-\x)   {};
      \node[inner sep=1pt] (UR\x) at (1,-\x+1) {};
      \node (MM\x) at ($(BL\x)!0.5!(UR\x)$) {};
      \draw (BL\x) rectangle (UR\x);
    }

    \node[anchor=west] at ($(MM0)-(0.2,0)$)  {\itshape TROJAN};
    \node[anchor=east] at ($(MM0)-(0.51,0)$) {0};
    \node[anchor=west] at ($(MM1)-(0.2,0)$)  {\ldots};
    \node[anchor=east] at ($(MM1)-(0.51,0)$) {\ldots};
    \node[anchor=west] at ($(MM2)-(0.2,0)$)  {\itshape TROJAN};
    \node[anchor=east] at ($(MM2)-(0.51,0)$) {$s-1$};
    \node[anchor=west] at ($(MM3)-(0.2,0)$)  {\itshape SPY};
    \node[anchor=east] at ($(MM3)-(0.51,0)$) {$s$};
    \node[anchor=west] at ($(MM4)-(0.2,0)$)  {\ldots};
    \node[anchor=east] at ($(MM4)-(0.51,0)$) {\ldots};
    \node[anchor=west] at ($(MM5)-(0.2,0)$)  {\itshape SPY};
    \node[anchor=east] (nsets) at ($(MM5)-(0.51,0)$) {$N - 1$};


    \draw ($(BL0|-UR0)+(0.25,0)$) -- ($(BL5)+(0.25,0)$);

    \node (state2left) at ($(BL2)-(0.5,0)$) {};
  \end{scope}

  \draw [-stealth] (state1right) -- node[above] {Encode} (state2left);

\end{tikzpicture}}
  \fi
  \vspace{3pt}
	\caption{A cache before and after the Trojan encodes the secret $s$.}
	\label{fig:prime}
\end{figure}
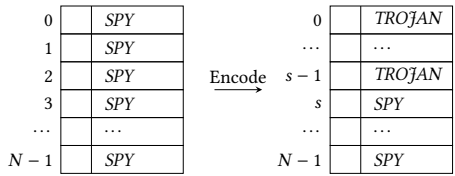

In a P\&P attack, the spy first forces the exploited hardware resource into a known state (\emph{prime}). For the D-cache this means traversing a large buffer (in cache-line-sized strides for efficiency), for the I-cache by executing a series of linked jumps. The TLBs are similarly primed by accessing or jumping with page-size strides. (This is a somewhat simplified description -- in general it is necessary to randomise the access order to prevent interference from prefetching, but that is not an issue on our processor.)
With a correctly-sized priming buffer, this leaves the hardware resource in a state where further accesses by the spy within the same address range are fast, as illustrated on the left of \autoref{fig:prime}.

At the end of its time slice, the OS preempts the spy and switches to the Trojan, which accesses a subset of the hardware resource to encode the secret. Given a D-cache of \(n\) lines, the Trojan can transmit a secret \(s\leq n\), the \emph{input signal}, by touching \(s\) cache lines, thereby replacing the spy's content. The resulting state is illustrated on the right of \autoref{fig:prime}. Obviously, more complex encodings are possible to increase the amount of data transferred in a time slice (the channel capacity), but for our purposes, the simple encoding is sufficient, as we want to prevent \emph{any} leakage.

When execution switches back to the spy, it again traverses (\emph{probes}) the whole buffer, observing its execution time. Each entry replaced by the Trojan's execution leads to a cache miss, and results in an increase in probe time. If the latency of a hit is \(t_{\textrm{hit}}\) and that of a miss is \(t_{\textrm{miss}} > t_{\textrm{hit}}\), the total latency increase is \(s \cdot (t_{\textrm{miss}} - t_{\textrm{hit}}).\) 
For our simple encoding scheme, the \emph{output signal} is the total probe time, which is linearly correlated to the input signal. A more sophisticated encoding scheme would have to measure the time of each individual access and perform a more complex analysis.

\subsection{Time Protection}

Time protection is a recently proposed, principled approach to  \emph{eliminating} timing channels~\cite{Ge2019a}. While the established notion of \emph{memory protection} prevents interference between security domains through unauthorised memory accesses, time protection aims to prevent interference that affects observable timing behaviour.

Time protection requires that all shared hardware resources, including non-architected ones, must be partitioned between security domains, either temporally (secure time multiplexing) or spatially. \citeauthor{Ge2019a} show that (physically-addressed) off-core caches can be effectively partitioned through \emph{cache colouring}~\cite{kessler1992colouring}, which leverages the associative cache lookup to force different partitions into disjoint subsets of the cache. They demonstrate that colouring is effective in
preventing cache channels in both intra-core and cross-core attacks and comes with low overhead.

Spatial partitioning is generally impossible for on-core resources for lack of hardware support. These are usually also fairly small and highly utilised by a single program, so partitioning would result in unacceptable performance degradation. Furthermore, on-core resources are accessed by virtual address, which is not under OS control, making approaches such as colouring infeasible.

This leaves temporal partitioning for on-core resources. In order to prevent any interference between security domains, each such resource must be reset to a state that is independent of execution history before handing it to a different domain. This means that the OS must be provided with the means to perform such a reset of all microarchitectural state, creating the requirement of extending the hardware-software contract to refer (in a highly abstract way) to such non-architected state~\cite{Ge2018}. The authors specifically show that contemporary Intel and Arm processors lack the mechanisms required for implementing time protection.

\subsection{Proposed Temporal Fence}\label{s:tfence}

We introduce such a mechanism in the form of a \emph{temporal fence} instruction, \code{fence.t}, which isolates the timing of any subsequent execution from what happened before.\footnote{Krste Asanovi\'c introduced the notion of a temporal fence on the RISC-V mailing list.}  Our fence instruction specifically applies to on-core state only, as off-core state can be spatially partitioned. We realise that this definition is less abstract than one might wish, and is therefore unlikely to be the last word on the topic. However, it suits our present purpose of evaluating the desired functionality.

We parameterise the \code{fence.t} instruction by the microarchitectural state targeted, as suggested by \citeauthor{Ge2018}. This helps the OS to minimise flushing according to its security requirements. For this study, it has the additional benefit that we can target individual channels for a fine-grained examination of efficacy.

\begin{figure}
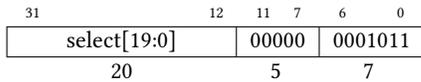
  \begin{tabular}{>{\centering\arraybackslash}p{20ex}>{\centering\arraybackslash}p{5ex}>{\centering\arraybackslash}p{7ex}}
    \instbitrange{31}{12} &
    \instbitrange{11}{7} &
    \instbitrange{6}{0}\\
    \hline
    \multicolumn{1}{|c|}{select[19:0]} &
    \multicolumn{1}{c|}{00000} &
    \multicolumn{1}{c|}{0001011}\\
    \hline
    20 & 5 & 7 \\
  \end{tabular}
  \caption{Encoding of the \code{fence.t} instruction.}
  \label{tab:encoding}
\end{figure}

We encode the \code{fence.t} instruction as a custom U-type instruction with the RISC-V opcode \textit{custom-0} (\autoref{tab:encoding}). A bitmap passed as the 20-bit immediate value selects the components to be flushed.

\section{Methodology}
\label{chap:methodology}

We adopt the approach of \citet{Ge2018} for quantifying and evaluation leakage and prevention strategies.



\subsection{Measuring Leakage}
\label{sec:metrics}

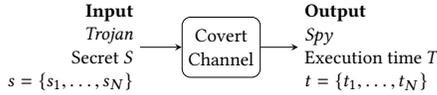
\begin{figure}
	\centering
        \ifFigsOn
	\scalebox{\figscale}{\def\NodeHeight{1cm}
\def\NodeDist{2.5cm}
\def\edgedist{2cm}

\begin{tikzpicture}
	
	\node (Input)   [node distance=\NodeDist, align=right] {
		\textbf{Input}\\
		\textit{Trojan}\\
		Secret $S$\\
		$s = \{s_1,\ldots,s_N\}$
	};
	\node (CovChan) [draw, rounded corners, minimum height=\NodeHeight, minimum width=1cm, node distance=\NodeDist, align=center, right of=Input] {Covert\\Channel};
	\node (Output)  [node distance=\NodeDist, align=left, right of=CovChan] {
	\textbf{Output}\\
	\textit{Spy}\\
	Execution time $T$\\
	$t = \{t_1,\ldots,t_N\}$
	};
	
	\draw [-stealth] (Input.east) -- (CovChan.west);
	\draw [-stealth] (CovChan.east) -- (Output.west);
	
\end{tikzpicture}}
        \fi 
	\caption{Relationship of the measured parameters.}
	\label{fig:information_theory}
\end{figure}

We run each attack for a number of iterations, the \emph{sample size}, usually 1~million. In each iteration, \(i\), the Trojan encodes as input value a randomly chosen secret, \(s_i\), and the spy subsequently measures as the output value its probe latency, \(t_i\). $s$ and $t$ can be regarded as samples of the random variables $S$ and $T$, see \autoref{fig:information_theory}. A covert channel exploits the correlation of the two random variables: If the output $t$ is correlated with the input $s$, there is a covert channel  that transfers information from the Trojan to the spy.

We use a combination of two indicators: The \emph{channel matrix} as a visual representation of leakage, and the \emph{discrete mutual information} $\mathcal{M}$ as a quantitative metric.

\subsubsection{Channel Matrix}
\label{subsec:channel_matrix}
The channel matrix represents the conditional probability of observing a particular output value, $t$, given input value $s$. The conditional probability distribution ${p(t\mid s)}$ can be computed directly from the measured sample pairs $\{(s_1, t_1), \ldots , (s_N, t_N)\}$.

We represent the channel matrix as a heat map: Inputs vary horizontally and outputs vertically, and bright colours indicate high, dark colours low probability (see \autoref{fig:l1d} for examples). A variation of colour along any horizontal line through the graph indicates a dependence of the output on the input, and thus a channel.


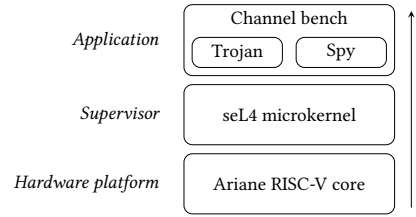
\begin{figure}
	\centering
        \ifFigsOn
	\scalebox{\figscale}{\def\NodeHeight{1cm}
\def\InnerNodeHeight{0.5cm}
\def\NodeWidth{3.5cm}
\def\InnerNodeWidth{1.5cm}
\def\NodeDist{1ex}
\def\edgedist{2cm}
\tikzstyle{Box} = [draw, rounded corners, minimum height=\NodeHeight, minimum width=\NodeWidth, node distance=\NodeDist, align=center]
\tikzstyle{InsideBox} = [draw, rounded corners, minimum height=\InnerNodeHeight, minimum width=\InnerNodeWidth, align=center]
\tikzstyle{Desc} = [node distance=0.3cm, align=right]

\begin{tikzpicture}[x=1ex,y=1ex]
	
	\node (ChanBench) [Box, text depth=0.75\NodeHeight] {Channel bench};
	\node[InsideBox, anchor=south west] (Trojan) at ($(ChanBench.south west)+(1,1)$) {Trojan};
	\node[InsideBox, anchor=south east] (Spy)    at ($(ChanBench.south east)+(-1,1)$) {Spy};
	\node (sel4) 	  [Box, below=of ChanBench] {seL4 microkernel};
	\node (Ariane)	  [Box, below=of sel4]      {Ariane RISC-V core};
	
	\node (Application) [Desc, left=of ChanBench] {\itshape Application};
	\node (Supervisor)	[Desc, left=of sel4]      {\itshape Supervisor};
	\node (HWPlat)		[Desc, left=of Ariane]	  {\itshape Hardware platform};
	
	\path (Ariane.south east)+(2,0) node(start) {};
	\path (ChanBench.north east)+(2,0) node(end) {};
	
	\draw [-stealth] (start) -- (end);
	
\end{tikzpicture}}
        \fi 
	\caption{Evaluation platform.}
	\label{fig:platform}
\end{figure}

\subsubsection{Mutual Information}
\label{subsec:mutual_information}

For quantifying channel capacity we use \emph{mutual information} $\mathcal{M}$, the amount of information gained about a random variable by observing another, possibly correlated random variable~\cite{Shannon_48}. It can be expressed as the difference between the marginal entropy $\textrm{H}(T)$ and the conditional entropy ${\textrm{H}(T \mid S)}$:
\[\mathcal{M} = \textrm{I}(S;T) = \textrm{H}(T) - \textrm{H}(T \mid S)\]

$\mathcal{M}$ is measured in bits; as most of our channel capacities are small, we use millibits ($1 \mathrm{\,mb}=10^{-3}\,\mathrm{b}$). Intuitively, mutual information is the difference of the information gained by observing the random variable $T$ \emph{without} and \emph{with} knowledge of the second random variable $S$. If both random variables are highly correlated (i.e., there exists a covert channel), the information gained by observing $S$ is low and the mutual information becomes high. Conversely, if both random variables are uncorrelated, we have $\textrm{H}(T) = \textrm{H}(T \mid S)$ and therefore $\mathcal{M} = 0$.

\paragraph{Zero Leakage Upper Bound $\mathcal{M}_0$}
Since all measurements are affected by noise, $\mathcal{M}$ will never be zero, even if there is no channel. We use a Monte Carlo simulation for estimating the apparent channel produced by this noise.  Specifically, we rearrange the input/output pairs into uniformly random pairs and thus remove any correlation between them, while retaining their original value ranges and spreads. Any mutual information that is measured from this data can only be due to noise. We repeat this process 1000 times and then compute the 95\%-confidence interval $\mathcal{M}_0$ for an experiment without a channel. We conclude that a channel is definitely present if $\mathcal{M} > \mathcal{M}_0$, else that the result is consistent with no channel.

We use the leakiEst tool~\cite{Chothia2013} to  compute mutual information $\mathcal{M}$ and zero leakage upper bounds $\mathcal{M}_0$. 


\subsection{Evaluation Platform}
\label{sec:platform}

\subsubsection{Ariane}
\label{subsec:ariane}

The hardware platform for evaluating  channels and defences is based on \textit{Ariane},  an open-source, RV64GC, 6-stage RISC-V core developed at ETH Zurich~\cite{Zaruba2019}. It is implemented in SystemVerilog and publicly available on GitHub~\cite{GitHub:Ariane}. It features three privilege levels and virtual memory (SV39) from the privileged ISA specification~\cite{Waterman2019}, and thus supports full-fledged operating systems. Its configurability, simplicity, and openness make it a good candidate for architectural exploration.

\paragraph{Setup}

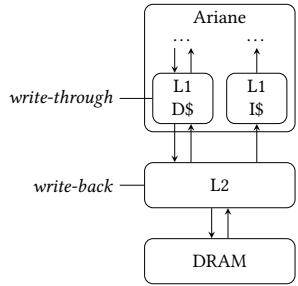
\begin{figure}
	\centering
        \ifFigsOn
	\scalebox{\figscale}{\def\NodeHeight{0.75cm}
\def\InnerNodeHeight{0.5cm}
\def\NodeWidth{2.5cm}
\def\InnerNodeWidth{1cm}
\def\NodeDist{0.5cm}
\def\edgedist{2cm}
\tikzstyle{Box} = [draw, rounded corners, minimum height=\NodeHeight, minimum width=\NodeWidth, node distance=\NodeDist, align=center]
\tikzstyle{InsideBox} = [draw, rounded corners, minimum height=\InnerNodeHeight, minimum width=\InnerNodeWidth, align=center]
\tikzstyle{Desc} = [node distance=0.3cm, align=right]

\begin{tikzpicture}[x=1ex,y=1ex]
	
	\node (ariane) [Box, text depth=1.7cm] {Ariane};
	\node[InsideBox, anchor=south west] (l1d) at ($(ariane.south west)+(1,1)$) {L1\\D\$};
	\node[InsideBox, anchor=south east] (l1i)    at ($(ariane.south east)+(-1,1)$) {L1\\I\$};
	\node (l1ddots) at ($(l1d.north)+(0,4)$) {\ldots};
	\node (l1idots) at ($(l1i.north)+(0,4)$) {\ldots};
	\node (l2) 	  [Box, below=of ariane] {L2};
	\node (dram)	  [Box, below=of l2]      {DRAM};
	\node[anchor=east] (wt) at ($(l1d.west)-(4,0)$) {\textit{write-through}};
	\node[anchor=east] (wb) at (wt.east|-l2.west) {\textit{write-back}};

	\path (l1d.south)+(-1,0) coordinate (l1dsource);
	\path (l1d.south)+(1,0)  coordinate (l1dsink);

	\draw [-stealth] (l1i.north) -- (l1idots.south);
	\draw [-stealth] ($(l1d.north)+(1,0)$) -- ($(l1ddots.south)+(1,0)$);
	\draw [-stealth] ($(l1ddots.south)-(1,0)$) -- ($(l1d.north)-(1,0)$);
	\draw [-stealth] (l1dsource) -- (l1dsource|-l2.north);
	\draw [-stealth] (l1dsink|-l2.north) -- (l1dsink);
	\draw [-stealth] (l1i.south|-l2.north) -- (l1i.south);
	\draw [-stealth] ($(dram.north)+(1,0)$) -- ($(l2.south)+(1,0)$);
	\draw [-stealth] ($(l2.south)-(1,0)$) -- ($(dram.north)-(1,0)$);
	\draw [-] (wt.east) -- (l1d.west);
	\draw [-] (wb.east) -- (l2.west);

	
\end{tikzpicture}}
        \fi 
        \vspace{3pt}
	\caption{Memory system of the Ariane SoC.}
	\label{fig:ariane-mem}
\end{figure}

We instantiate the Ariane core on an FPGA (Digilent Genesys II), running at 50~MHz, using the standard configuration with an 8-way, 32\,KiB write-through L1-D and a 4-way, 16\,KiB L1-I cache. Both use 16-byte lines and a pseudo-random replacement strategy driven by an 8-bit linear-feedback shift register (LFSR). The L1-D is accessed by the load-, store-, and memory-management units, with concurrent accesses arbitrated round-robin. The branch predictor has a 64-entry BHT and a 16-entry BTB. There is a single-level, unified, fully associative, 16-entry TLB using a pseudo-LRU replacement policy. For reducing write-stalls we increase the write buffer to 40 entries. We add some off-core components, including a timer and a 512\,KiB write-back L2 cache~\cite{wolfgang2019llc-thesis} that is connected to DRAM.  \autoref{fig:ariane-mem} shows the memory architecture.

We partition the L2 cache by colouring~\cite{kessler1992colouring}, which precludes channels in the memory backend and allows us to focus on channels resulting from on-core state.

\newlength\figH
\newlength\figW
\setlength{\figW}{0.8\linewidth}
\setlength{\figH}{0.618\figW} 

\pgfplotsset{every axis/.append style={
label style={font=\small},
tick label style={font=\small},
x tick label style={/pgf/number format/.cd, precision=1},
scaled y ticks = false,
y tick label style={/pgf/number format/.cd, fixed, precision=1},
xtick distance = 64,
colorbar style = {at={(1.05,0)}, anchor=south west}
}}

\newlength\subfigdist
\setlength{\subfigdist}{2ex}

\captionsetup[figure]{aboveskip=0ex}
\begin{figure*}[t]
    \captionsetup[subfigure]{aboveskip=-2ex,belowskip=1ex}
    \begin{subfigure}[t]{0.45\linewidth}
        \ifFigsOn
\begin{tikzpicture}

\definecolor{color0}{rgb}{0.267004,0.004874,0.329415}

\begin{axis}[
axis background/.style={fill=color0},
colorbar,
colorbar style={ytick={0,0.111848310923165,0.160854937272428,0.189521975973601,0.209861563621692,0.225638173479167,0.238528602322864,0.249427304618208,0.258868189970955,0.267195641024037,0.274644799828431,0.323651426177694,0.352318464878867},yticklabels={\(\displaystyle {0}\),\(\displaystyle {10^{-2}}\),,,,,,,,,\(\displaystyle {10^{-1}}\),,},ylabel={Probability}},
colormap/viridis,
height=\figH,
point meta max=0.3664678633213,
point meta min=0,
tick align=outside,
tick pos=left,
width=\figW,
x grid style={white!69.0196078431373!black},
xlabel={Secret},
xmin=0, xmax=256,
xtick style={color=black},
y grid style={white!69.0196078431373!black},
ylabel={Time (cycles)},
ymin=76021, ymax=89640,
ytick style={color=black}
]
\addplot graphics [includegraphics cmd=\pgfimage,xmin=0, xmax=256, ymin=76021, ymax=89640] {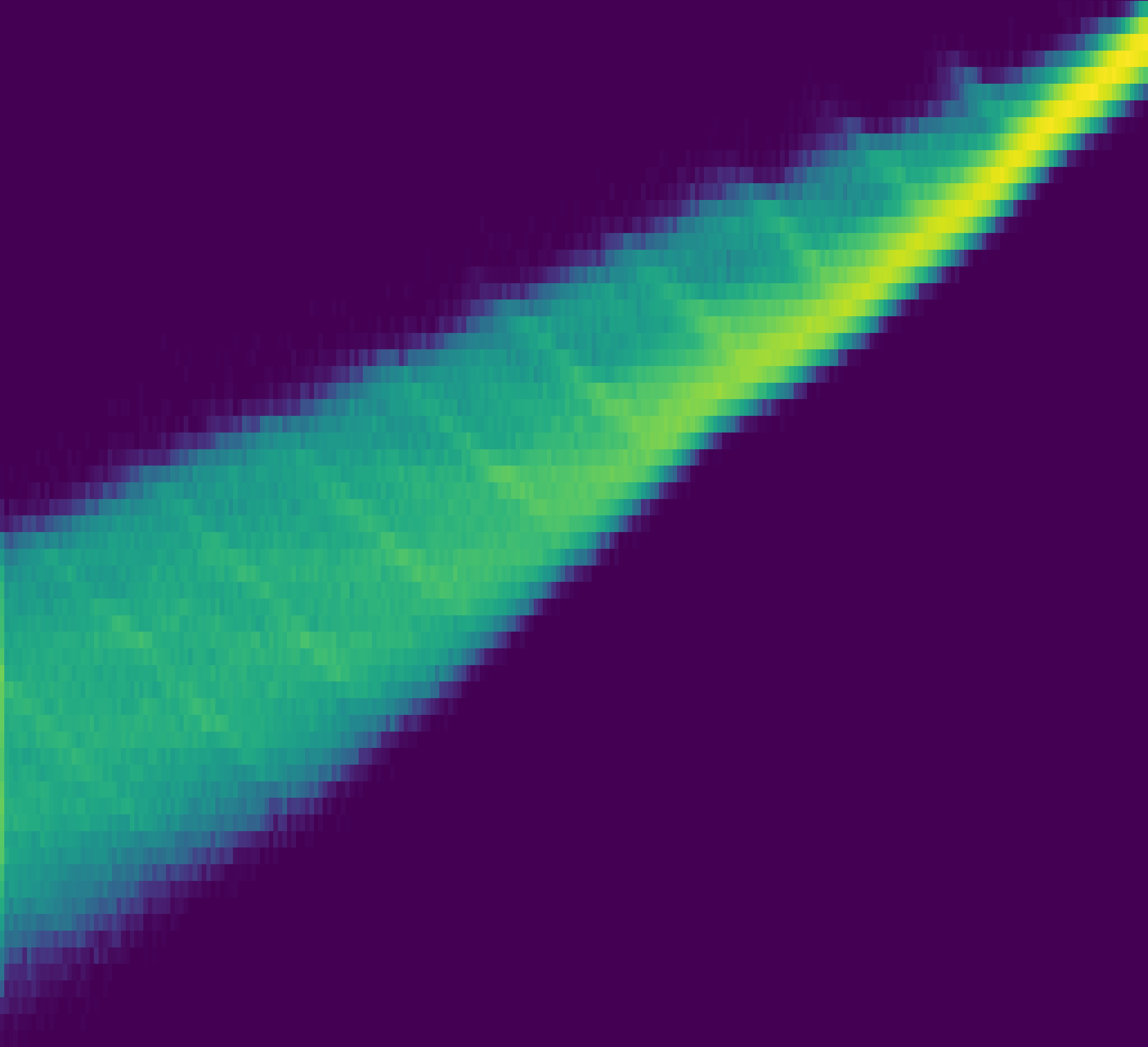};
\end{axis}

\end{tikzpicture}
        \fi 
        \caption{Unmitigated. \normalfont $N=10^6$, $\mathcal{M} = 1667.3\text{mb}$, $\mathcal{M}_0 = 0.5\text{mb}$.}
        \label{fig:l1d-unmitigated}
    \end{subfigure}
    \hfill
    \begin{subfigure}[t]{0.45\linewidth}
      \ifFigsOn
\begin{tikzpicture}

\definecolor{color0}{rgb}{0.267004,0.004874,0.329415}

\begin{axis}[
axis background/.style={fill=color0},
colorbar,
colorbar style={ytick={0,0.0818746264142459,0.117748205470123,0.138732903984467,0.153621784526001,0.16517049748825,0.174606483040345,0.182584495148586,0.189495363581878,0.195591181554689,0.201044076544128,0.236917655600005},yticklabels={\(\displaystyle {0}\),\(\displaystyle {10^{-2}}\),,,,,,,,,\(\displaystyle {10^{-1}}\),},ylabel={Probability}},
colormap/viridis,
height=\figH,
point meta max=0.2480640262365,
point meta min=0,
tick align=outside,
tick pos=left,
width=\figW,
x grid style={white!69.0196078431373!black},
xlabel={Secret},
xmin=0, xmax=256,
xtick style={color=black},
y grid style={white!69.0196078431373!black},
ylabel={Time (cycles)},
ymin=91795, ymax=93003,
ytick style={color=black}
]
\addplot graphics [includegraphics cmd=\pgfimage,xmin=0, xmax=256, ymin=91795, ymax=93003] {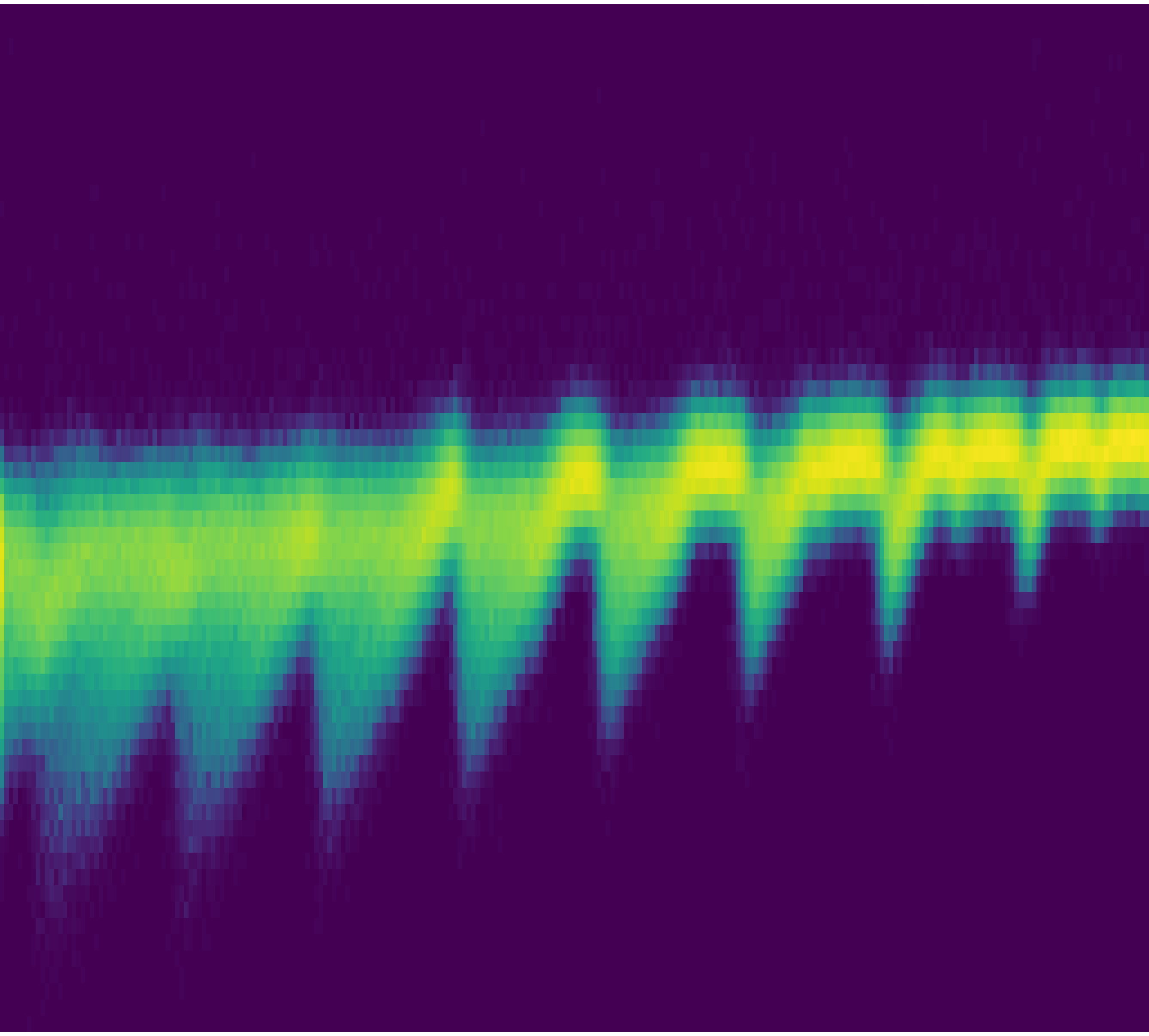};
\end{axis}

\end{tikzpicture}
        \fi 
        \caption{Attempted reset by priming twice. \normalfont $N=10^6$, $\mathcal{M} = 515.7\text{mb}$, $\mathcal{M}_0 = 1.1\text{mb}$.}
        \label{fig:l1d-sw}
    \end{subfigure}
    \begin{subfigure}[t]{0.45\linewidth}
      \ifFigsOn
\begin{tikzpicture}

\definecolor{color0}{rgb}{0.267004,0.004874,0.329415}

\begin{axis}[
axis background/.style={fill=color0},
colorbar,
colorbar style={ytick={0,0.137352136230329,0.197533329517521,0.23273707083918,0.257714522804713,0.277088539707707,0.292918264126372,0.306302105420427,0.317895716091905,0.328122005448031,0.337269732994899,0.39745092628209,0.432654667603749,0.457632119569282,0.477006136472276,0.492835860890941,0.506219702184997,0.517813312856474,0.5280396022126,0.537187329759468,0.59736852304666,0.632572264368319,0.657549716333852,0.676923733236846,0.692753457655511,0.706137298949566},yticklabels={\(\displaystyle {0}\),\(\displaystyle {10^{-3}}\),,,,,,,,,\(\displaystyle {10^{-2}}\),,,,,,,,,\(\displaystyle {10^{-1}}\),,,,,,},ylabel={Probability}},
colormap/viridis,
height=\figH,
point meta max=0.707001388073,
point meta min=0,
tick align=outside,
tick pos=left,
width=\figW,
x grid style={white!69.0196078431373!black},
xlabel={Secret},
xmin=0, xmax=256,
xtick style={color=black},
y grid style={white!69.0196078431373!black},
ylabel={Time (cycles)},
ymin=92507, ymax=92581,
ytick style={color=black}
]
\addplot graphics [includegraphics cmd=\pgfimage,xmin=0, xmax=256, ymin=92507, ymax=92581] {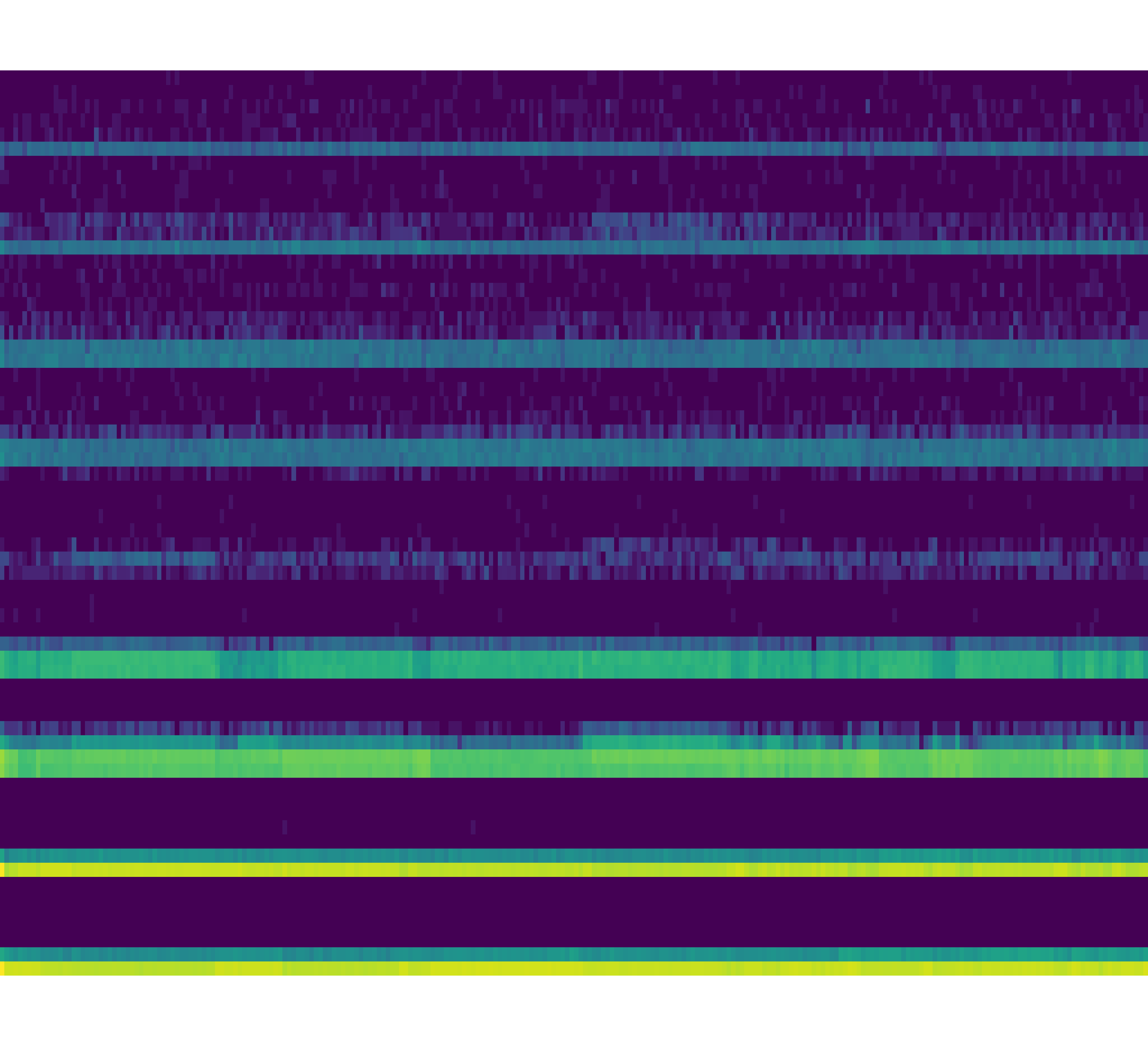};
\end{axis}

\end{tikzpicture}
        \fi 
        \caption{Original fence clearing first-order state. \normalfont $N=10^6$, ${\mathcal{M} = 10.4\text{mb}}$, ${\mathcal{M}_0 = 6.0\text{mb}}$.}
        \label{fig:l1d-1st-order}
    \end{subfigure}
    \hfill
    \begin{subfigure}[t]{0.45\linewidth}
      \ifFigsOn
\begin{tikzpicture}

\definecolor{color0}{rgb}{0.267004,0.004874,0.329415}

\begin{axis}[
axis background/.style={fill=color0},
colorbar,
colorbar style={ytick={0,0.1064647343211,0.153112532675712,0.180399745454363,0.199760331030325,0.214777567885315,0.227047543808975,0.237421660635072,0.246408129384937,0.254334756587625,0.261425366239927,0.30807316459454,0.33536037737319,0.354720962949152,0.369738199804142,0.382008175727802,0.392382292553899,0.401368761303764,0.409295388506453,0.416385998158755,0.463033796513367,0.490321009292017,0.509681594867979,0.524698831722969,0.53696880764663,0.547342924472726,0.556329393222592,0.56425602042528,0.571346630077582,0.617994428432194,0.645281641210845,0.664642226786807,0.679659463641797,0.691929439565457,0.702303556391554},yticklabels={\(\displaystyle {0}\),\(\displaystyle {10^{-4}}\),,,,,,,,,\(\displaystyle {10^{-3}}\),,,,,,,,,\(\displaystyle {10^{-2}}\),,,,,,,,,\(\displaystyle {10^{-1}}\),,,,,,},ylabel={Probability}},
colormap/viridis,
height=\figH,
point meta max=0.7025480866432,
point meta min=0,
tick align=outside,
tick pos=left,
width=\figW,
x grid style={white!69.0196078431373!black},
xlabel={Secret},
xmin=0, xmax=257,
xtick style={color=black},
y grid style={white!69.0196078431373!black},
ylabel={Time (cycles)},
ymin=92527, ymax=92626,
ytick style={color=black}
]
\addplot graphics [includegraphics cmd=\pgfimage,xmin=0, xmax=257, ymin=92527, ymax=92626] {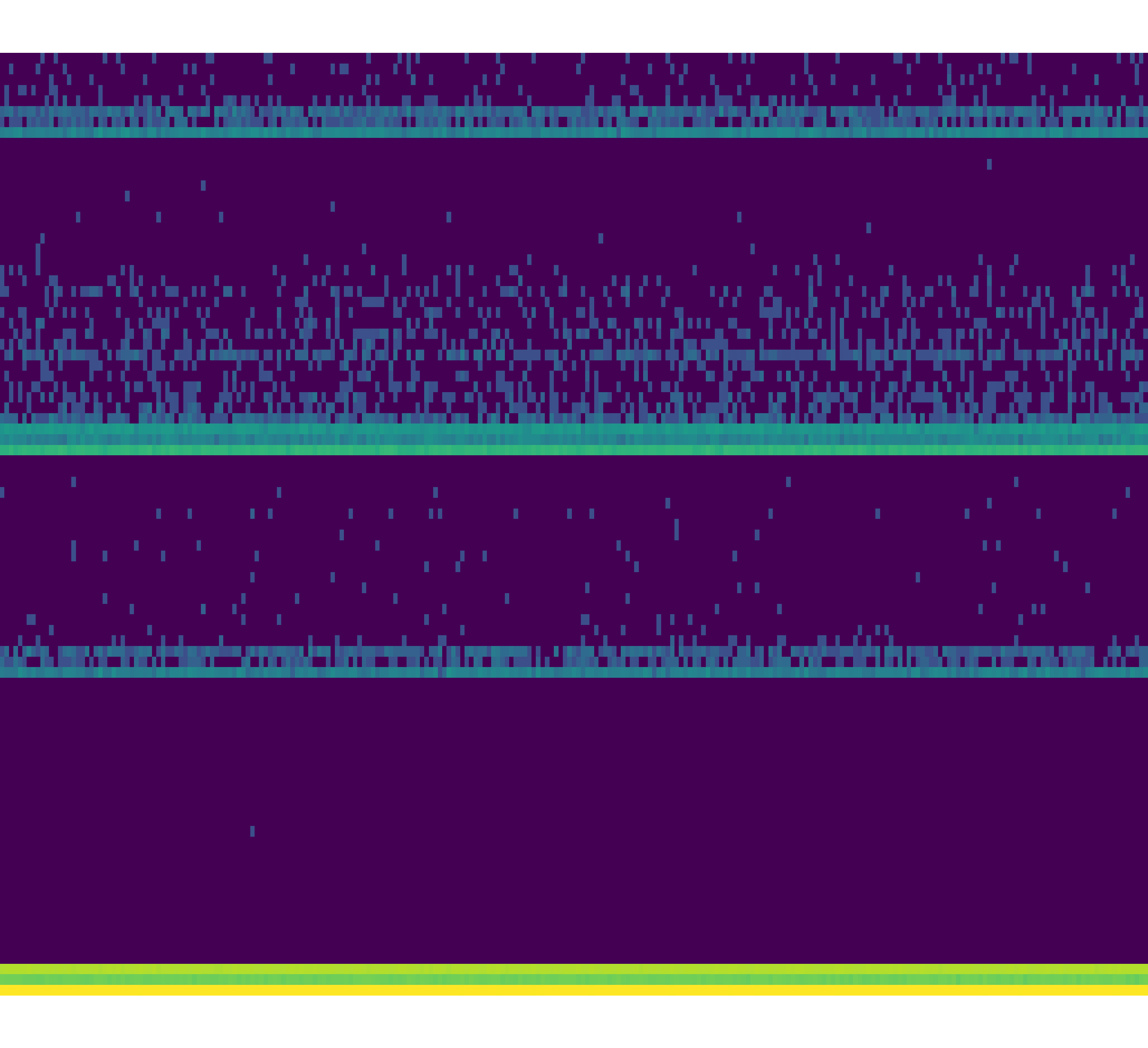};
\end{axis}

\end{tikzpicture}
        \fi 
        \caption{Improved fence clearing first- and second-order state. \normalfont $N=10^6$, $\mathcal{M} = 33.3\text{mb}$, $\mathcal{M}_0 = 39.1\text{mb}$.}
        \label{fig:l1d-fencet}
    \end{subfigure}
    \caption{L1 data cache channel matrices.}
    \label{fig:l1d}
\end{figure*}

\subsubsection{OS}
\label{subsec:sel4}\label{sec:attacks}
Ge's \emph{channel bench}~\cite{Ge2019phd,GitHub:channel-bench} provides a minimal OS and data collection infrastructure; we port it to RISC-V and adapt to Ariane. Channel bench uses attack implementations from the \emph{Mastik} toolkit~\cite{Yarom2016}, running on an experimental version of seL4~\cite{Klein2014_seL4} that supports time protection. seL4 is an open-source, high-performance OS microkernel with formal proofs of implementation correctness and security enforcement, making it highly suitable for security evaluations, although our experimental version is not verified.


\section{Covert-Channel Capacities}

\subsection{Baseline: No Time Protection}

To establish a baseline and compare to other architectures, we apply the P\&P attacks to our Ariane RV64GC core, with time protection disabled in seL4. We observe strong channels through each of the five microarchitectural resources targeted. As shown in the \emph{Unmitigated} column of \autoref{tab:mutinf}, capacities range from 0.4 to 4\,bits. The $\mathcal{M}_0$ are all well below 1\,mb, indicating that the channels are real. To put those numbers into context: Assuming the OS uses a 1\,ms time slice, Trojan and spy will each execute 500 times per second. The 1.6\,b capacity of the D-cache thus means the channel has a bandwidth of 833\,b/s, able to leak a 1024-bit RSA key in just over a second. Also, these channels use a rather primitive encoding scheme;  more sophistication could increase the bandwidth significantly.

The channel capacities we observe agree nicely with the prior work, which showed unmitigated capacities of 0.3--4\,b on Intel and 7.5\,mb to 2.5\,b on Arm processors~\cite{Ge2019a}.

\autoref{fig:l1d-unmitigated} and \autoref{fig:bht-unmitigated} show the unmitigated channel matrices for the L1-D cache and the BHT, respectively; \(N\) is the number of iterations. The clear diagonal pattern indicates a strong correlation of output with input signals, establishing efficient channels. For example, \autoref{fig:bht-unmitigated} shows that if the spy observes a probe time of 380 cycles, it can infer with a high confidence that the Trojan has encoded the value \emph{48}.

\begin{table}
    \caption[Mutual information and corresponding zero leakage upper bound]{Mutual information and corresponding zero leakage upper bound in millibits.}
    \label{tab:mutinf}
    \begin{tabular}{%
        l
        c
        S[table-format=4.1]
        S[table-format=1.1]
        c
        S[table-format=2.1]
        S[table-format=2.1]
        c
        S[table-format=2.1]
        S[table-format=2.1]}
        \toprule
        && \multicolumn{2}{c}{Unmitigated} && \multicolumn{2}{c}{First \code{fence.t}} && \multicolumn{2}{c}{Final \code{fence.t}} \\
        \cmidrule{3-4} \cmidrule{6-7} \cmidrule{9-10}
        && ${\mathcal{M}}$ & ${\mathcal{M}_0}$ && ${\mathcal{M}}$ & ${\mathcal{M}_0}$ && ${\mathcal{M}}$ & ${\mathcal{M}_0}$ \\
        \midrule
        L1 D\$ && 1667.3 & 0.5 && 10.4 &  6.0 && 33.3 & 39.1 \\
        L1 I\$ && 1905.0 & 0.5 &&  8.3 &  4.9 && 37.9 & 39.4 \\
        TLB    &&  408.7 & 0.1 &&  5.0 &  5.9 &&  3.1 &  7.7 \\
        BTB    && 3211.4 & 0.1 && 35.7 & 59.3 && 28.2 & 60.3 \\
        BHT    && 3770.6 & 0.2 && 45.2 & 58.8 && 44.1 & 60.8 \\
        \bottomrule
    \end{tabular}
\end{table}


\pgfplotsset{every axis/.append style={xtick distance=16}}
\begin{figure*}[t]
    \captionsetup[subfigure]{aboveskip=-2ex,belowskip=1ex}
    \begin{subfigure}[t]{0.45\linewidth}
        \ifFigsOn
\begin{tikzpicture}

\definecolor{color0}{rgb}{0.267004,0.004874,0.329415}

\begin{axis}[
axis background/.style={fill=color0},
colorbar,
colorbar style={ytick={0,0.146253260204356,0.210334503953492,0.247819628546308,0.274415747702628,0.2950453004208,0.311900872295445,0.326152055255066,0.338496991451764,0.349385996888261,0.359126544169936,0.423207787919073,0.460692912511889,0.487289031668209,0.507918584386381,0.524774156261025,0.539025339220647,0.551370275417345,0.562259280853842,0.571999828135517,0.636081071884653,0.67356619647747,0.70016231563379,0.720791868351962,0.737647440226606,0.751898623186228,0.764243559382926,0.775132564819423,0.784873112101098,0.848954355850234,0.886439480443051,0.91303559959937,0.933665152317543,0.950520724192187,0.964771907151808,0.977116843348507,0.988005848785004},yticklabels={\(\displaystyle {0}\),\(\displaystyle {10^{-4}}\),,,,,,,,,\(\displaystyle {10^{-3}}\),,,,,,,,,\(\displaystyle {10^{-2}}\),,,,,,,,,\(\displaystyle {10^{-1}}\),,,,,,,,},ylabel={Probability}},
colormap/viridis,
height=\figH,
point meta max=0.9975165128708,
point meta min=0,
tick align=outside,
tick pos=left,
width=\figW,
x grid style={white!69.0196078431373!black},
xlabel={Secret},
xmin=0, xmax=65,
xtick style={color=black},
y grid style={white!69.0196078431373!black},
ylabel={Time (cycles)},
ymin=162, ymax=488,
ytick style={color=black}
]
\addplot graphics [includegraphics cmd=\pgfimage,xmin=0, xmax=65, ymin=162, ymax=488] {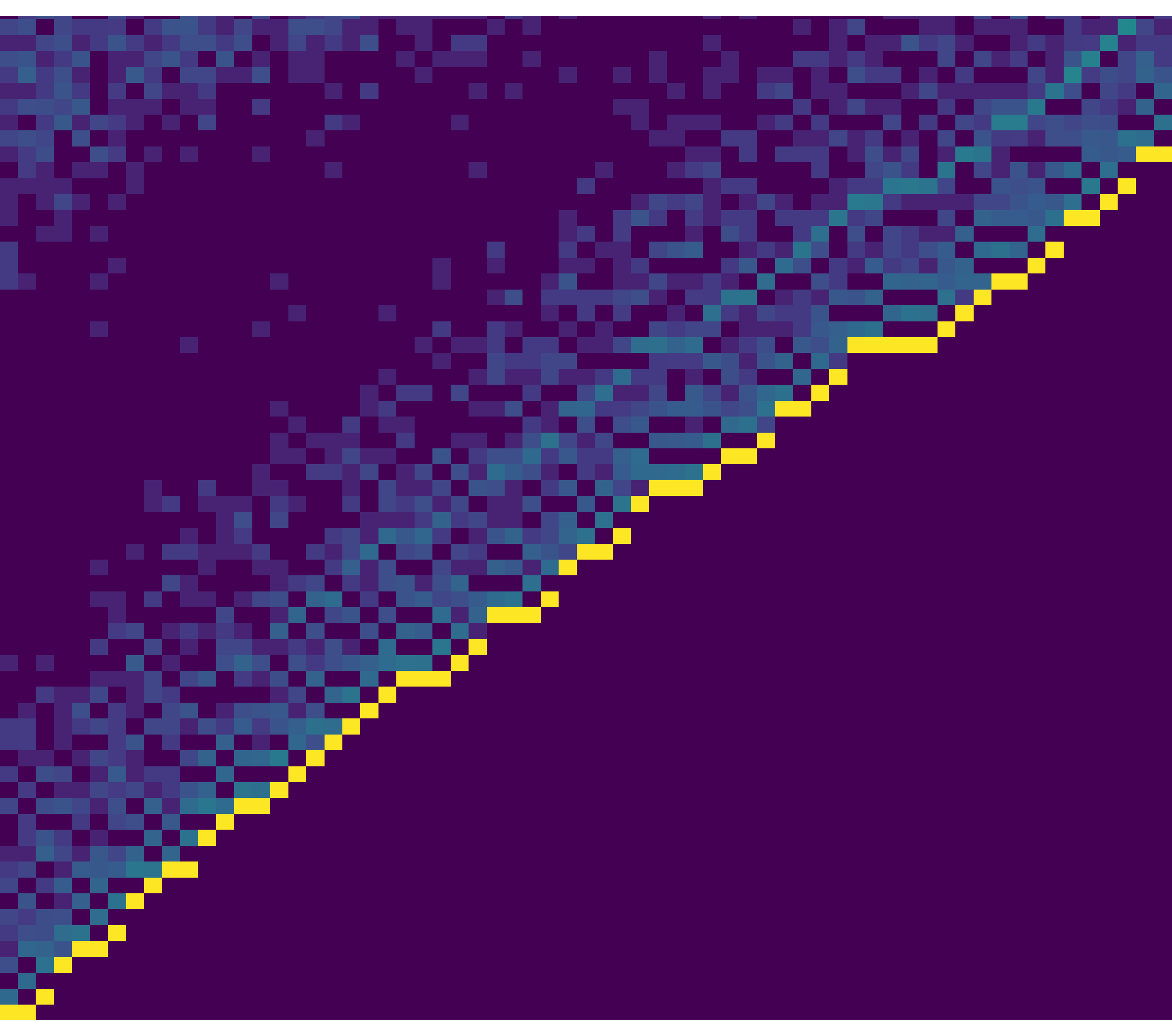};
\draw[->,very thick,draw=red] (0, 380) -- (47, 380);
\draw[->,very thick,draw=red] (48, 380) -- (48, 170);
\end{axis}

\end{tikzpicture}
        \fi 
        \caption{Unmitigated. \normalfont $N=10^6$, $\mathcal{M} = 3770.6\text{mb}$, $\mathcal{M}_0 = 0.2\text{mb}$.}
        \label{fig:bht-unmitigated}
    \end{subfigure}
    \hfill
    \begin{subfigure}[t]{0.45\linewidth}
        \ifFigsOn
\begin{tikzpicture}

\definecolor{color0}{rgb}{0.267004,0.004874,0.329415}

\begin{axis}[
axis background/.style={fill=color0},
colorbar,
colorbar style={ytick={0,0.120721976450847,0.173616622272861,0.204558006567666,0.226511268094875,0.243539540654098,0.25745265238968,0.269216020749772,0.279405913916889,0.288394036684485,0.296434186476112,0.349328832298125,0.380270216592931,0.402223478120139,0.419251750679362,0.433164862414944,0.444928230775036,0.455118123942153,0.46410624670975,0.472146396501376,0.52504104232339,0.555982426618195,0.577935688145404,0.594963960704626,0.608877072440209,0.6206404408003,0.630830333967417,0.639818456735014,0.64785860652664,0.700753252348654,0.731694636643459,0.753647898170668,0.770676170729891,0.784589282465473,0.796352650825565,0.806542543992682,0.815530666760278,0.823570816551905,0.876465462373918,0.907406846668724,0.929360108195932,0.946388380755155,0.960301492490737,0.972064860850829,0.982254754017946,0.991242876785542},yticklabels={\(\displaystyle {0}\),\(\displaystyle {10^{-5}}\),,,,,,,,,\(\displaystyle {10^{-4}}\),,,,,,,,,\(\displaystyle {10^{-3}}\),,,,,,,,,\(\displaystyle {10^{-2}}\),,,,,,,,,\(\displaystyle {10^{-1}}\),,,,,,,,},ylabel={Probability}},
colormap/viridis,
height=\figH,
point meta max=0.9992237687111,
point meta min=0,
tick align=outside,
tick pos=left,
width=\figW,
x grid style={white!69.0196078431373!black},
xlabel={Secret},
xmin=0, xmax=65,
xtick style={color=black},
y grid style={white!69.0196078431373!black},
ylabel={Time (cycles)},
ymin=1197, ymax=1271,
ytick style={color=black}
]
\addplot graphics [includegraphics cmd=\pgfimage,xmin=0, xmax=65, ymin=1197, ymax=1271] {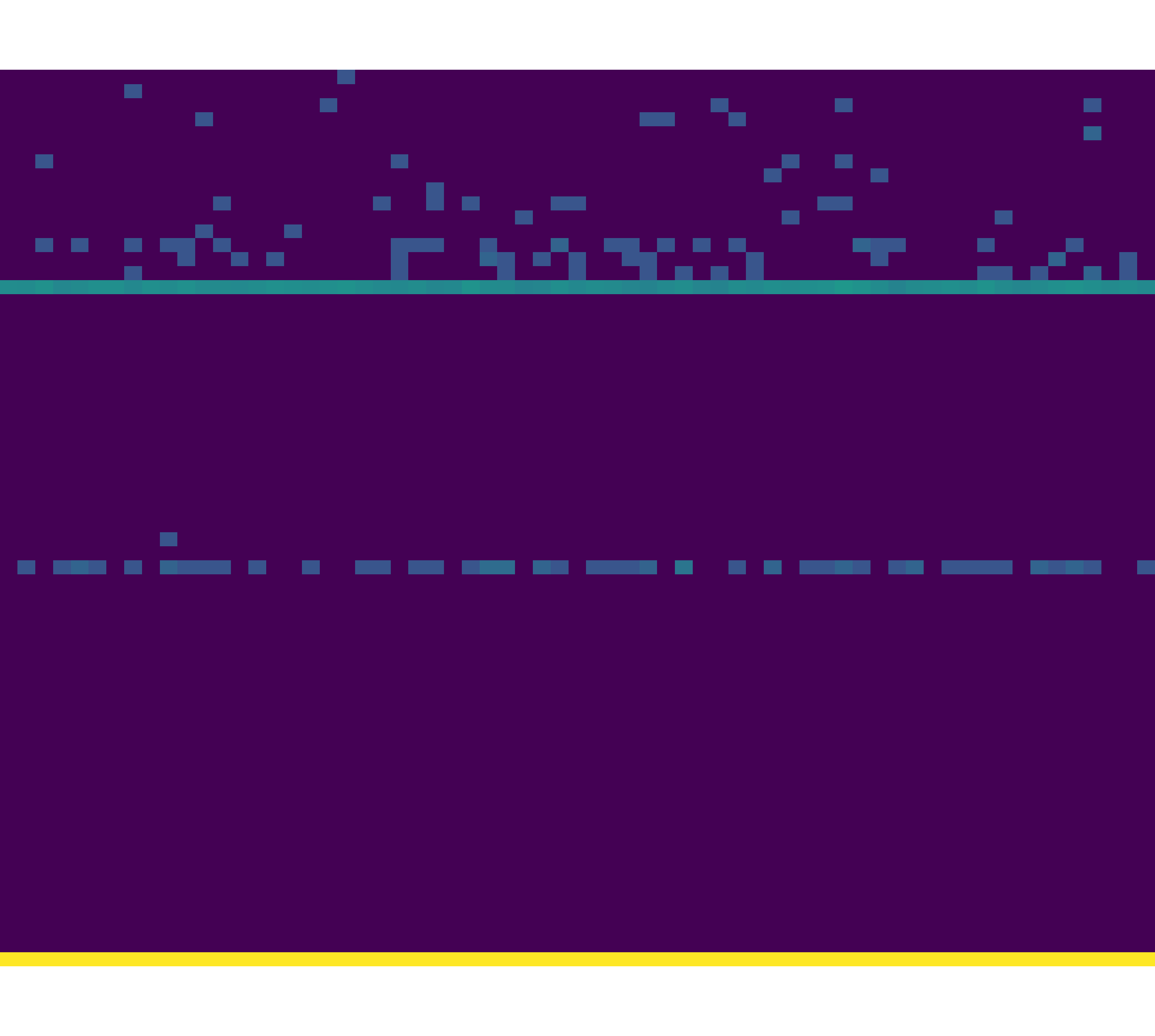};
\end{axis}

\end{tikzpicture}
        \fi 
        \caption{Improved fence. \normalfont $N=10^6$, $\mathcal{M} = 44.1\text{mb}$, $\mathcal{M}_0 = 60.8\text{mb}$.}
        \label{fig:bht-fencet}
    \end{subfigure}
    \caption{BHT channel matrices.}
    \label{fig:bht}
\end{figure*}

\subsection{Using Existing Instructions Only}\label{s:swflush}

\citeauthor{Ge2018} report that neither the x86 nor the Arm architectures provide sufficient mechanisms for implementing time protection, with Arm coming closer in at least providing targeted L1 cache flushes. The Intel architecture does not have those, and the authors implemented software flushing by touching all cache lines, similar to the prime phase of the P\&P attack. Such an approach is expensive and obviously brittle, as it must make assumptions on the replacement policy which may not hold in reality. Unsurprisingly, they find that this defence is incomplete, leaving residual channels that the OS is unable to close.

With RISC-V, the situation is presently worse, as specification of cache management is still under discussion. While implementations generally support some cache management, this is consequently not standardised. To explore this aspect, we implement a ``software only'' defence, where the OS uses only mechanisms defined in the ISA as presently specified. This basically forces the OS to resort to the priming approach in an attempt to erase any microarchitectural state left by the Trojan's execution.

\autoref{fig:l1d-sw} shows the result for the L1-D cache channel, where the OS performs \emph{two} priming runs per context switch. While fuzzier than in the unmitigated case, a clear diagonal pattern persists, and the measured capacity is only reduced by 70\%, making this defence highly ineffective, a result that is much worse than \citeauthor{Ge2018} observed for Intel. The reasons are for one the Ariane's replacement policy, which uses a pseudo-random sequence with a period of 256. This makes it practically impossible to flush the cache reliably through priming. Furthermore, there is secondary state that is even harder to reset, as we will find in \autoref{subsubsec:2nd-degree}.


\subsection{Using a Temporal Fence}
\label{subsubsec:fence-t}

As discussed in \autoref{s:tfence} we add a new \code{fence.t} instruction to the Ariane. When \code{fence.t} is committed, Ariane's controller sends a flush signal to all stateful microarchitectural components. To give the operating system maximum control, an immediate value selects the components to be flushed.

\subsubsection{First Attempt}
\label{subsubsec:1st-degree}

As we want to minimise the cost of the fence, we only flush state that seems to be relevant for the P\&P attacks. In particular, we reset the L1 cache and TLB state by clearing the valid bits (remember, our Ariane's L1-D is a write-through cache so there is no dirty state to write back). Similarly, we reset the saturation counter of the BHT. To avoid interfering with in-flight computations, we also flush the pipeline.

We find that this is insufficient, as shown in \autoref{fig:l1d-1st-order}. While the channel pattern is gone, the channel is not quite closed: The channel matrix shows slight patters along horizontal lines. The mutual information, shown in \autoref{tab:mutinf} as ``First \code{fence.t}'',  is almost twice the zero-leakage bound, confirming the channel.

\subsubsection{Improved Fence}
\label{subsubsec:2nd-degree}

Investigating the source, we identify further state that indirectly affects timing. Most prominently,
\begin{itemize}
	\item the LFSR used for victim selection in L1 cache replacement
	\item the round robin arbiter of the L1 data cache
	\item the pseudo-LRU tree for the TLB replacement strategy.
\end{itemize}

We include these components in the flush and re-run the experiments,  the results are shown as ``Final \texttt{fence.t}'' in \autoref{tab:mutinf}. All measured channel capacities are now clearly below the zero-capacity threshold, meaning that there is no evidence of a residual channel. The channel matrices confirm this:  \autoref{fig:l1d-fencet} and \autoref{fig:bht-fencet} only show patterns that appear to be random noise (confirmed by visual comparison between multiple runs). 


\section{Cost}

\subsection{Context-Switch Latency}
\label{subsec:context-switch}

Time protection resets hardware state on a switch of security partition, which implies a full context switch. seL4's IPC is essentially a user-triggered context switch~\cite{Heiser_Elphinstone_16} with roughly the same cost as a time-slice preemption, and the seL4 benchmark suite~\cite{github:sel4bench} provides a convenient rig for measuring its latency. We use inter-address-space IPC for evaluating flush cost. \autoref{tab:context-switch} compares the latencies of various configurations. Here ``hot'' measures the best-case of switching for and back in a tight loop, where the whole working set fits into the L1 caches. The cold-cache scenario is the realistic baseline for our purposes, as a security-domain switch is normally triggered by time-slice preemption; as time slices are 1\,ms or longer, the newly executing domain is unlikely to have any hot data left in the small L1 caches. We achieve the cold state by executing \code{fence.t} from user mode (before the timed context-switch).

The third column shows the latency with the OS trying to reset state by double priming as discussed in \autoref{s:swflush}, note that this only attempts to mitigate the D-cache channel. Finally, ``\code{fence.t}'' uses the full flush provided by the temporal fence.

\begin{table}
\caption{seL4 IPC latencies and standard deviations in cycles.}
\label{tab:context-switch}
\begin{tabular}{ccccc}
    \toprule
    \multicolumn{2}{c}{Unmitigated} && \multicolumn{2}{c}{Mitigated} \\
    \cmidrule{1-2} \cmidrule{4-5}
    Hot & Cold && D-cache prime & \texttt{fence.t} \\
    \midrule
    430 ($\pm$7.0) & 1,180 ($\pm$1.0) && 51,877 ($\pm$256) & 1,502 ($\pm$0.9) \\
    \bottomrule
\end{tabular}
\end{table}

We already found in \autoref{s:swflush} that the software priming is highly ineffective, the results here show that it is also very expensive, increasing context-switch latency by a factor of 50 over the cold-cache case (while not even attempting to mitigate channels other than the L1-D). In contrast, the temporal fence, which we have found to be highly effective against \emph{all} channels, only adds 320 cycles (less than 30\%) to the cold-cache latency. With a switch rate of no more than 1\,kHz, this adds negligible cost. 

The dominating contribution to the direct latency of the \texttt{fence.t} instruction is the cache flush. A write-through cache is flushed by clearing all valid bits. This is a constant-time operation, which could in theory be performed in a single cycle. However, in Ariane's write-through cache, the valid bits are stored together with the tags in sequentially accessible SRAM, allowing for an invalidation of only one set per cycle, and thus resulting in a latency of 256 cycles. All other state can be reset in a single cycle.

A write-back L1-D cache would be more expensive to flush, as each dirty line must we written back to the L2. Since the L2 of our platform can process up to 8\,B per cycle, the theoretical latency for a write-back variant varies between 0 cycles (clean cache) and 4,096 cycles (all lines dirty). In such a case of a variable latency, the OS must pad to the worst-case latency, to prevent the flush latency becoming a covert channel~\cite{Ge2018}.

\subsection{Hardware Overhead}

To estimate the hardware costs incurred by the \texttt{fence.t} instruction, we examine the resource utilization of our FPGA. The number of deployed LUTs remains within 1\% of the original size. Hence, the mechanism should not cause a notable increase in chip area or power draw of the design.



\section{Related Work}
Past work has approached the hardware mitigation of microarchitectural covert channels from different angles.
\citet{Page_05} propose static partitioning of the L1 cache while \citet{wang2007cache-mapping1} propose locking cache lines. While spatial partitioning can certainly prevent attacks, in the case of the L1, the reduction of available cache space would have a major impact on application performance. \citet{wang2008cache-mapping2} instead aim to defeat attacks by dynamically remapping cache lines.

A hardware feature that aims to detect an ongoing cache-based covert channel attack is proposed by \citet{chen2014cc-hunter}.
Fang et al.\ present a method to scramble information transmitted over such a channel by leveraging cache prefetchers~\cite{fang2018prefetch-guard,fang2019prodact}. This is not applicable to Ariane, which lacks prefetching. 

\citet{Fadiheh2019} suggest a formal method for detecting vulnerable microarchitectural components within the HW design. While such an approach could prove crucial for the systematic uncovering of microarchitectural covert channels, the question of their mitigation remains open.

Our work extends that of Ge et al., who propose time protection and the need for flushing all microarchitectural on-core state on a partition switch, and demonstrate the need for hardware support~\cite{Ge2018,Ge2019a,Ge2019phd}, which is what our temporal fence provides.



\section{Conclusion}

On a simple in-order application-class RISC-V processor we evaluate microarchitectural covert timing channels, previously demonstrated on x86 and Arm processors, and find that they exist with similar capacities on the RISC-V core. We confirm the finding of \citet{Ge_YCH_18} that existing architected mechanisms are insufficient for preventing those channels. Answering their request for improved hardware support that will enable a principled prevention of such channels, we propose a temporal fence instruction, \code{fence.t}.

An implementation of \code{fence.t} on our RISC-V core shows that the naive approach of just clearing all valid bits on cache lines is insufficient. Instead we find that secondary state, in our case the state machine controlling cache-line replacement, can also be exploited as a covert channel, and must be reset as well. We then demonstrate that a complete state flush is successful in eliminating all channels to well below measurement accuracy. We also find that while the (largely ineffective) attempts to close channels with existing instructions are extremely costly, the overhead of \code{fence.t} is very low, about 320~cycles on our core, which is insignificant at typical partition-switch rates of 1\,kHz or lower. We similarly find that the area and power overheads of \code{fence.t} are insignificant.

Our findings show that the mechanisms requested by OS researchers for principled timing-channel prevention are feasible and low cost, and there seems to be no good reason not to include them into the architecture. This confirms that security should be seen as a hardware-software codesign problem, where OS researchers and architects must collaborate closely.

We hope our findings will support current work that aims at provably eliminating microarchitectural timing channels~\cite{Heiser_KM_19}.

\begin{acks}
We thank Qian Ge and Curtis Millar for their support with the covert channel measurement framework, Wolfgang R\"onninger for providing us with an L2 cache, and Florian Zaruba for his help and insights on Ariane. The support of HENSOLDT Cyber and the IDEA League for Wistoff's work at ETH is gratefully acknowledged. Heiser's work was supported by Australian Research Council (ARC) grant DP190103743 and the US Asian Office of Aerospace Research and Development (AOARD).
\end{acks}

\bibliographystyle{ACM-Reference-Format}
\bibliography{sources}

%
\appendix

\ifFigsOn
\onecolumn
\section{Channel Matrices}
\label{app:channel-matrices}

\pgfplotsset{every axis/.append style={
label style={font=\small},
tick label style={font=\small},
x tick label style={/pgf/number format/.cd, precision=1},
scaled y ticks = false,
y tick label style={/pgf/number format/.cd, fixed, precision=1},
xtick distance = 64,
colorbar style = {at={(1.05,0)}, anchor=south west}
}}

\captionsetup[figure]{aboveskip=0ex,belowskip=1ex}

\begin{figure*}[h!]
    \captionsetup[subfigure]{aboveskip=-2ex,belowskip=1ex}
    \begin{subfigure}[t]{0.45\linewidth}
\begin{tikzpicture}

\definecolor{color0}{rgb}{0.267004,0.004874,0.329415}

\begin{axis}[
axis background/.style={fill=color0},
colorbar,
colorbar style={ytick={0,0.111848310923165,0.160854937272428,0.189521975973601,0.209861563621692,0.225638173479167,0.238528602322864,0.249427304618208,0.258868189970955,0.267195641024037,0.274644799828431,0.323651426177694,0.352318464878867},yticklabels={\(\displaystyle {0}\),\(\displaystyle {10^{-2}}\),,,,,,,,,\(\displaystyle {10^{-1}}\),,},ylabel={Probability}},
colormap/viridis,
height=\figH,
point meta max=0.3664678633213,
point meta min=0,
tick align=outside,
tick pos=left,
width=\figW,
x grid style={white!69.0196078431373!black},
xlabel={Secret},
xmin=0, xmax=256,
xtick style={color=black},
y grid style={white!69.0196078431373!black},
ylabel={Time (cycles)},
ymin=76021, ymax=89640,
ytick style={color=black}
]
\addplot graphics [includegraphics cmd=\pgfimage,xmin=0, xmax=256, ymin=76021, ymax=89640] {figures/l1d_sm-000.png};
\end{axis}

\end{tikzpicture}
        \caption{Unmitigated. \normalfont $N=10^6$, $\mathcal{M} = 1667.3\text{mb}$, $\mathcal{M}_0 = 0.5\text{mb}$.}
    \end{subfigure}
    \hfill
    \begin{subfigure}[t]{0.45\linewidth}
\begin{tikzpicture}

\definecolor{color0}{rgb}{0.267004,0.004874,0.329415}

\begin{axis}[
axis background/.style={fill=color0},
colorbar,
colorbar style={ytick={0,0.0716000621749738,0.10297181436892,0.121323112984296,0.134343566562867,0.144443014979942,0.152694865178242,0.159671705110221,0.165715318756813,0.171046163793617,0.175814767173889,0.207186519367835},yticklabels={\(\displaystyle {0}\),\(\displaystyle {10^{-2}}\),,,,,,,,,\(\displaystyle {10^{-1}}\),},ylabel={Probability}},
colormap/viridis,
height=\figH,
point meta max=0.2092280834913,
point meta min=0,
tick align=outside,
tick pos=left,
width=\figW,
x grid style={white!69.0196078431373!black},
xlabel={Secret},
xmin=0, xmax=256,
xtick style={color=black},
y grid style={white!69.0196078431373!black},
ylabel={Time (cycles)},
ymin=90531, ymax=92558,
ytick style={color=black}
]
\addplot graphics [includegraphics cmd=\pgfimage,xmin=0, xmax=256, ymin=90531, ymax=92558] {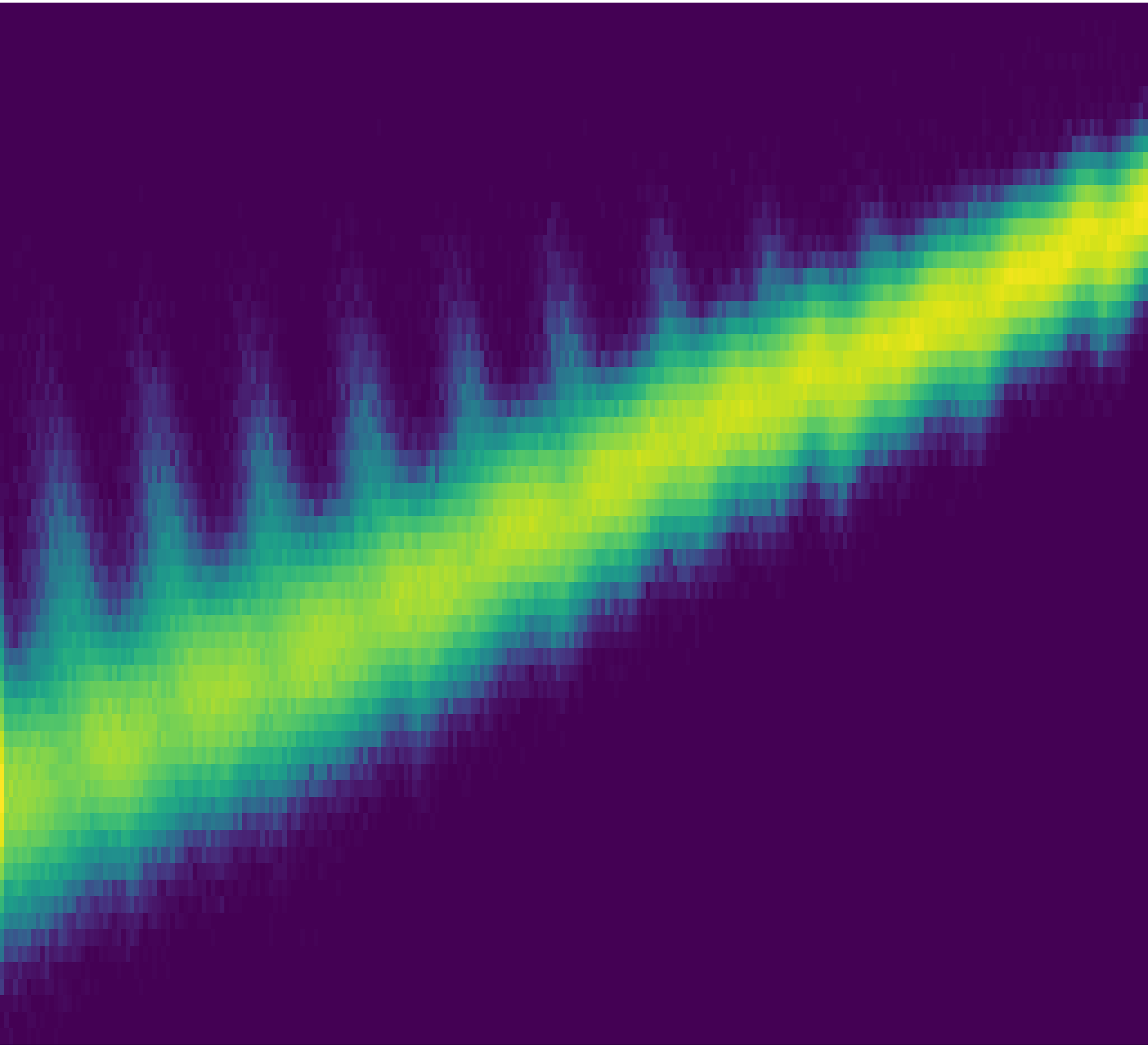};
\end{axis}

\end{tikzpicture}
        \caption{Attempted reset by priming once. \normalfont $N=10^6$, $\mathcal{M} = 1471.5\text{mb}$, $\mathcal{M}_0 = 0.5\text{mb}$.}
    \end{subfigure}
    \begin{subfigure}[t]{0.45\linewidth}
\begin{tikzpicture}

\definecolor{color0}{rgb}{0.267004,0.004874,0.329415}

\begin{axis}[
axis background/.style={fill=color0},
colorbar,
colorbar style={ytick={0,0.0818746264142459,0.117748205470123,0.138732903984467,0.153621784526001,0.16517049748825,0.174606483040345,0.182584495148586,0.189495363581878,0.195591181554689,0.201044076544128,0.236917655600005},yticklabels={\(\displaystyle {0}\),\(\displaystyle {10^{-2}}\),,,,,,,,,\(\displaystyle {10^{-1}}\),},ylabel={Probability}},
colormap/viridis,
height=\figH,
point meta max=0.2480640262365,
point meta min=0,
tick align=outside,
tick pos=left,
width=\figW,
x grid style={white!69.0196078431373!black},
xlabel={Secret},
xmin=0, xmax=256,
xtick style={color=black},
y grid style={white!69.0196078431373!black},
ylabel={Time (cycles)},
ymin=91795, ymax=93003,
ytick style={color=black}
]
\addplot graphics [includegraphics cmd=\pgfimage,xmin=0, xmax=256, ymin=91795, ymax=93003] {figures/l1d_sw2_sm-000.png};
\end{axis}

\end{tikzpicture}
        \caption{Attempted reset by priming twice. \normalfont $N=10^6$, $\mathcal{M} = 515.7\text{mb}$, $\mathcal{M}_0 = 1.1\text{mb}$.}
    \end{subfigure}
    \hfill
    \begin{subfigure}[t]{0.45\linewidth}
\begin{tikzpicture}

\definecolor{color0}{rgb}{0.267004,0.004874,0.329415}

\begin{axis}[
axis background/.style={fill=color0},
colorbar,
colorbar style={ytick={0,0.137352136230329,0.197533329517521,0.23273707083918,0.257714522804713,0.277088539707707,0.292918264126372,0.306302105420427,0.317895716091905,0.328122005448031,0.337269732994899,0.39745092628209,0.432654667603749,0.457632119569282,0.477006136472276,0.492835860890941,0.506219702184997,0.517813312856474,0.5280396022126,0.537187329759468,0.59736852304666,0.632572264368319,0.657549716333852,0.676923733236846,0.692753457655511,0.706137298949566},yticklabels={\(\displaystyle {0}\),\(\displaystyle {10^{-3}}\),,,,,,,,,\(\displaystyle {10^{-2}}\),,,,,,,,,\(\displaystyle {10^{-1}}\),,,,,,},ylabel={Probability}},
colormap/viridis,
height=\figH,
point meta max=0.707001388073,
point meta min=0,
tick align=outside,
tick pos=left,
width=\figW,
x grid style={white!69.0196078431373!black},
xlabel={Secret},
xmin=0, xmax=256,
xtick style={color=black},
y grid style={white!69.0196078431373!black},
ylabel={Time (cycles)},
ymin=92507, ymax=92581,
ytick style={color=black}
]
\addplot graphics [includegraphics cmd=\pgfimage,xmin=0, xmax=256, ymin=92507, ymax=92581] {figures/l1d_1st-order_sm-000.png};
\end{axis}

\end{tikzpicture}
        \caption{Original fence clearing first-order state. \normalfont $N=10^6$, ${\mathcal{M} = 10.4\text{mb}}$, ${\mathcal{M}_0 = 6.0\text{mb}}$.}
    \end{subfigure}
    \begin{subfigure}[t]{0.45\linewidth}
\begin{tikzpicture}

\definecolor{color0}{rgb}{0.267004,0.004874,0.329415}

\begin{axis}[
axis background/.style={fill=color0},
colorbar,
colorbar style={ytick={0,0.214079291947847,0.307878686708579,0.362747815233948,0.401678081469311,0.431874741926222,0.45654720999468,0.477407484515413,0.495477476230044,0.511416338520049,0.525674136686954,0.619473531447686,0.674342659973055,0.713272926208419,0.743469586665329,0.768142054733788,0.78900232925452,0.807072320969151,0.823011183259157,0.837268981426061,0.931068376186794,0.985937504712163,1.02486777094753,1.05506443140444,1.07973689947289,1.10059717399363,1.11866716570826,1.13460602799826,1.14886382616517,1.2426632209259,1.29753234945127,1.33646261568663,1.36665927614354,1.391331744212,1.41219201873273,1.43026201044737,1.44620087273737,1.46045867090428},yticklabels={\(\displaystyle {0}\),\(\displaystyle {10^{-4}}\),,,,,,,,,\(\displaystyle {10^{-3}}\),,,,,,,,,\(\displaystyle {10^{-2}}\),,,,,,,,,\(\displaystyle {10^{-1}}\),,,,,,,,,\(\displaystyle {10^{0}}\)},ylabel={Probability}},
colormap/viridis,
height=\figH,
point meta max=1.516838431358,
point meta min=0,
tick align=outside,
tick pos=left,
width=\figW,
x grid style={white!69.0196078431373!black},
xlabel={Secret},
xmin=0, xmax=256,
xtick style={color=black},
y grid style={white!69.0196078431373!black},
ylabel={Time (cycles)},
ymin=92527, ymax=92601,
ytick style={color=black}
]
\addplot graphics [includegraphics cmd=\pgfimage,xmin=0, xmax=256, ymin=92527, ymax=92601] {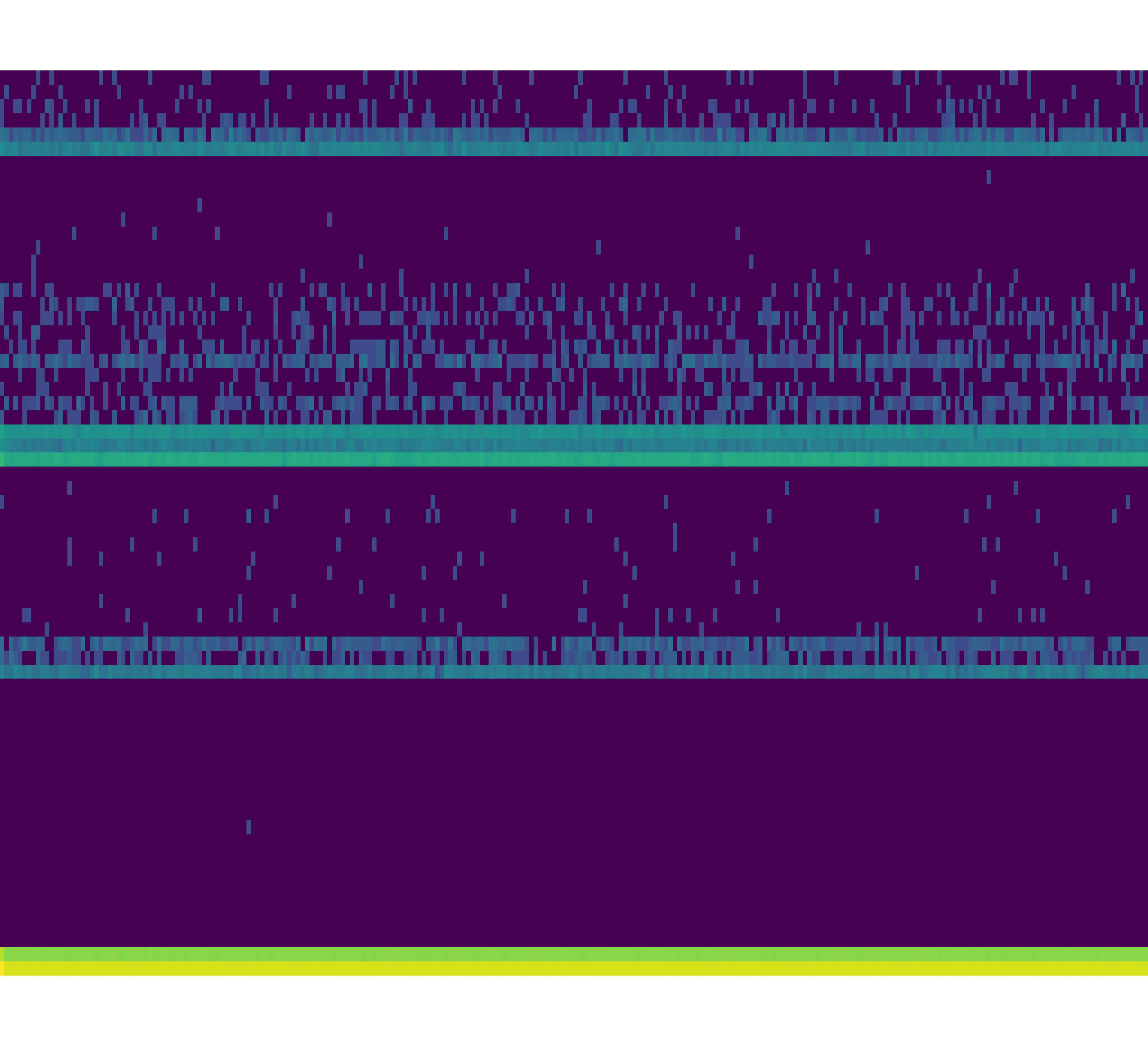};
\end{axis}

\end{tikzpicture}
        \caption{Improved fence (1). \normalfont $N=10^6$, $\mathcal{M} = 33.3\text{mb}$, $\mathcal{M}_0 = 39.1\text{mb}$.}
    \end{subfigure}
    \hfill
    \begin{subfigure}[t]{0.45\linewidth}
\begin{tikzpicture}

\definecolor{color0}{rgb}{0.267004,0.004874,0.329415}

\begin{axis}[
axis background/.style={fill=color0},
colorbar,
colorbar style={ytick={0,0.21650000174734,0.311360036759719,0.366849600059056,0.406220071772097,0.436758182124581,0.461709635071435,0.482805787945898,0.501080106784476,0.517199198370772,0.53161821713696,0.626478252149339,0.681967815448676,0.721338287161717,0.751876397514201,0.776827850461055,0.797924003335518,0.816198322174096,0.832317413760392,0.84673643252658,0.941596467538958,0.997086030838296,1.03645650255134,1.06699461290382,1.09194606585067,1.11304221872514,1.13131653756372,1.14743562915001,1.1618546479162,1.25671468292858,1.31220424622792,1.35157471794096,1.38211282829344,1.40706428124029,1.42816043411476,1.44643475295334,1.46255384453963,1.47697286330582},yticklabels={\(\displaystyle {0}\),\(\displaystyle {10^{-4}}\),,,,,,,,,\(\displaystyle {10^{-3}}\),,,,,,,,,\(\displaystyle {10^{-2}}\),,,,,,,,,\(\displaystyle {10^{-1}}\),,,,,,,,,\(\displaystyle {10^{0}}\)},ylabel={Probability}},
colormap/viridis,
height=\figH,
point meta max=1.535679578781,
point meta min=0,
tick align=outside,
tick pos=left,
width=\figW,
x grid style={white!69.0196078431373!black},
xlabel={Secret},
xmin=0, xmax=256,
xtick style={color=black},
y grid style={white!69.0196078431373!black},
ylabel={Time (cycles)},
ymin=92517, ymax=92591,
ytick style={color=black}
]
\addplot graphics [includegraphics cmd=\pgfimage,xmin=0, xmax=256, ymin=92517, ymax=92591] {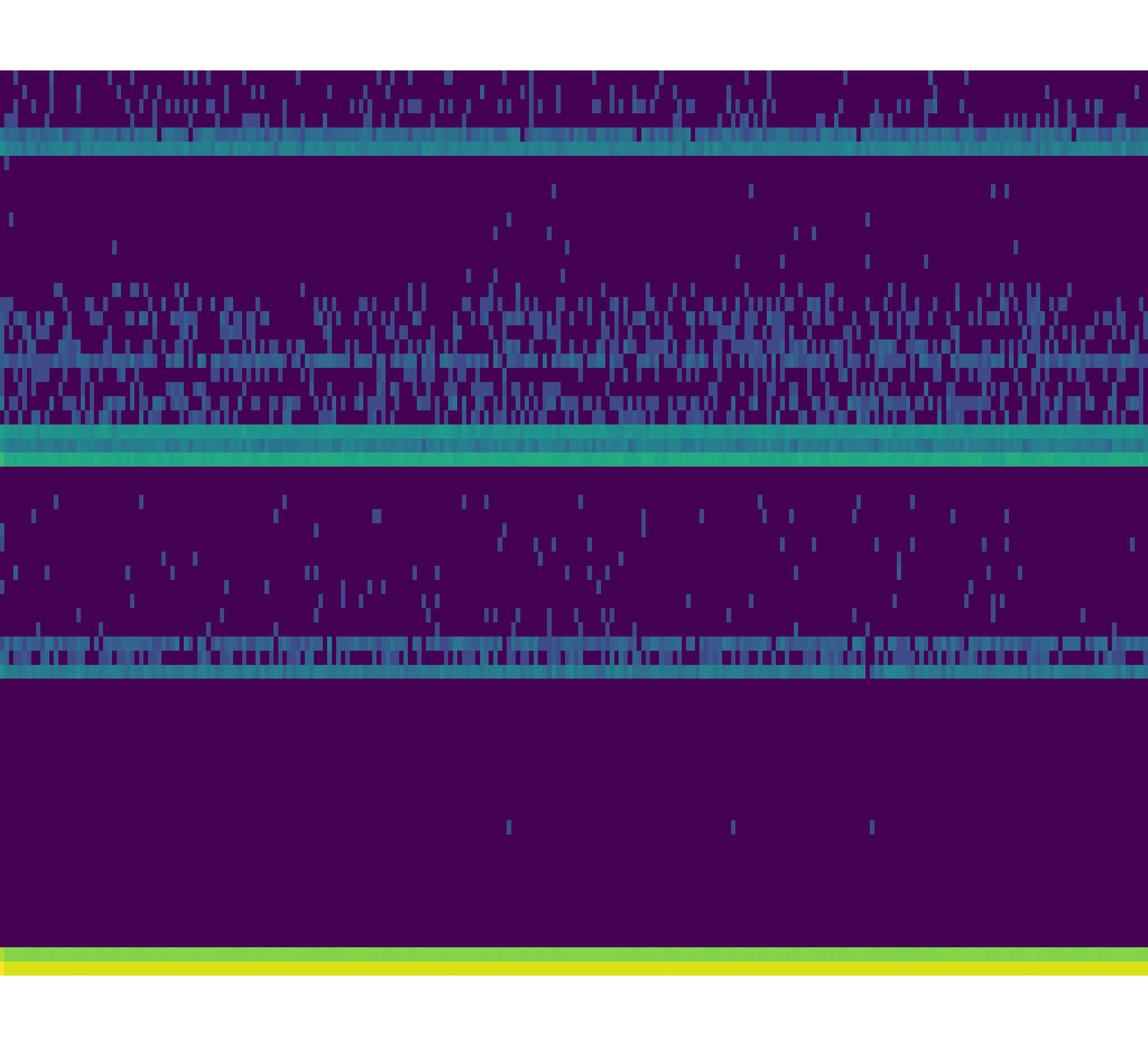};
\end{axis}

\end{tikzpicture}
        \caption{Improved fence (2). \normalfont $N=10^6$, $\mathcal{M} = 31.1\text{mb}$, $\mathcal{M}_0 = 38.6\text{mb}$.}
    \end{subfigure}
    \caption{L1 data cache channel matrices.}
    \label{fig:l1d-full}
\end{figure*}

\begin{figure*}[h]
    \captionsetup[subfigure]{aboveskip=-2ex,belowskip=1ex}
    \begin{subfigure}[t]{0.45\linewidth}
\begin{tikzpicture}

\definecolor{color0}{rgb}{0.267004,0.004874,0.329415}

\begin{axis}[
axis background/.style={fill=color0},
colorbar,
colorbar style={ytick={0,0.104435095756018,0.150193602730527,0.176960613399603,0.195952109705037,0.210683058680231,0.222719120374112,0.232895465536914,0.241710616679546,0.249486131043188,0.25644156565474,0.30220007262925,0.328967083298325},yticklabels={\(\displaystyle {0}\),\(\displaystyle {10^{-2}}\),,,,,,,,,\(\displaystyle {10^{-1}}\),,},ylabel={Probability}},
colormap/viridis,
height=\figH,
point meta max=0.3365581035614,
point meta min=0,
tick align=outside,
tick pos=left,
width=\figW,
x grid style={white!69.0196078431373!black},
xlabel={Secret},
xmin=0, xmax=256,
xtick style={color=black},
y grid style={white!69.0196078431373!black},
ylabel={Time (cycles)},
ymin=19071, ymax=24500,
ytick style={color=black}
]
\addplot graphics [includegraphics cmd=\pgfimage,xmin=0, xmax=256, ymin=19071, ymax=24500] {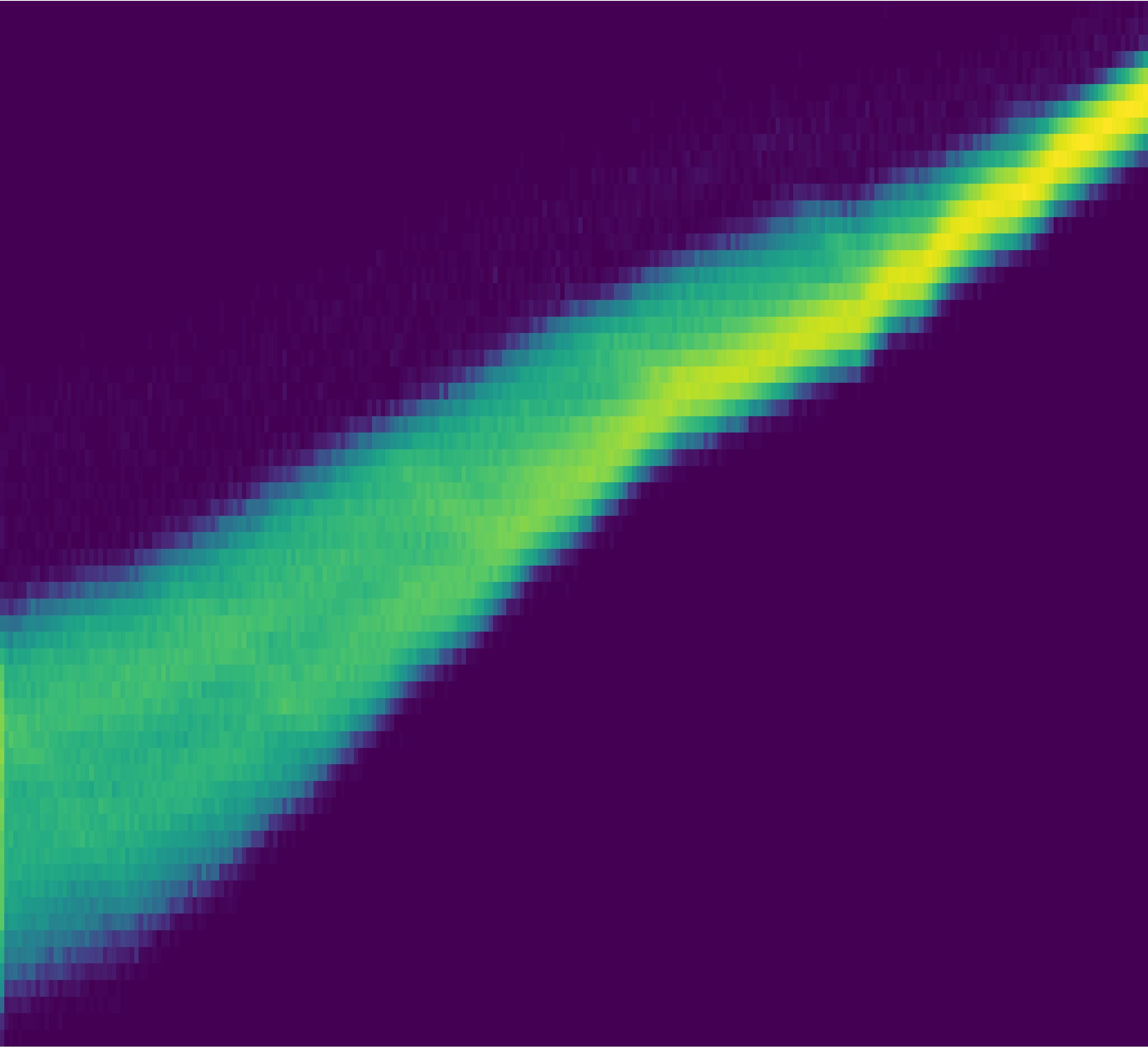};
\end{axis}

\end{tikzpicture}
        \caption{Unmitigated. \normalfont $N=10^6$, $\mathcal{M} = 1905.0\text{mb}$, $\mathcal{M}_0 = 0.5\text{mb}$.}
    \end{subfigure}
    \hfill
    \begin{subfigure}[t]{0.45\linewidth}
\begin{tikzpicture}

\definecolor{color0}{rgb}{0.267004,0.004874,0.329415}

\begin{axis}[
axis background/.style={fill=color0},
colorbar,
colorbar style={ytick={0,0.271233662408572,0.390075392223611,0.459593347686244,0.50891712203865,0.547175613711124,0.578435077501283,0.604864577550662,0.627758851853689,0.647953032963916,0.666017343526163,0.784859073341202,0.854377028803835,0.903700803156241,0.941959294828715,0.973218758618874,0.999648258668253,1.02254253297128,1.04273671408151,1.06080102464375,1.17964275445879,1.24916070992143,1.29848448427383,1.33674297594631,1.36800243973647,1.39443193978584,1.41732621408887,1.4375203951991,1.45558470576135,1.57442643557638,1.64394439103902,1.69326816539142,1.7315266570639,1.76278612085406,1.78921562090344,1.81210989520646,1.83230407631669,1.85036838687894},yticklabels={\(\displaystyle {0}\),\(\displaystyle {10^{-4}}\),,,,,,,,,\(\displaystyle {10^{-3}}\),,,,,,,,,\(\displaystyle {10^{-2}}\),,,,,,,,,\(\displaystyle {10^{-1}}\),,,,,,,,,\(\displaystyle {10^{0}}\)},ylabel={Probability}},
colormap/viridis,
height=\figH,
point meta max=1.966296195984,
point meta min=0,
tick align=outside,
tick pos=left,
width=\figW,
x grid style={white!69.0196078431373!black},
xlabel={Secret},
xmin=0, xmax=256,
xtick style={color=black},
y grid style={white!69.0196078431373!black},
ylabel={Time (cycles)},
ymin=25367, ymax=25441,
ytick style={color=black}
]
\addplot graphics [includegraphics cmd=\pgfimage,xmin=0, xmax=256, ymin=25367, ymax=25441] {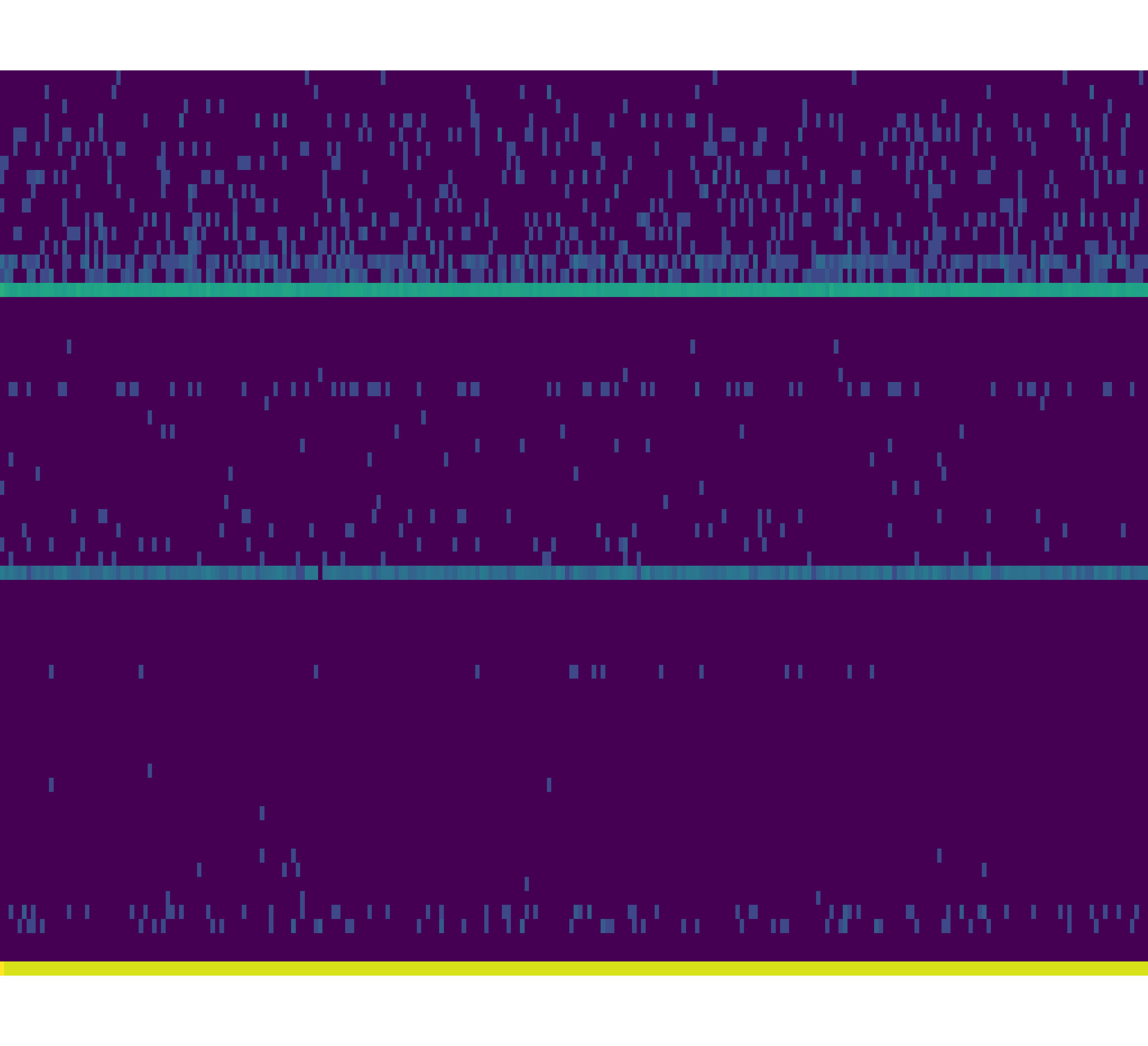};
\end{axis}

\end{tikzpicture}
        \caption{Improved fence. \normalfont $N=10^6$, $\mathcal{M} = 37.9\text{mb}$, $\mathcal{M}_0 = 39.4\text{mb}$.}
    \end{subfigure}
    \caption{L1 instruction cache channel matrices.}
    \label{fig:l1i}
\end{figure*}

\pgfplotsset{every axis/.append style={xtick distance=4}}
\begin{figure*}
    \captionsetup[subfigure]{aboveskip=-2ex,belowskip=1ex}
    \begin{subfigure}[t]{0.45\linewidth}
\begin{tikzpicture}

\definecolor{color0}{rgb}{0.267004,0.004874,0.329415}

\begin{axis}[
axis background/.style={fill=color0},
colorbar,
colorbar style={ytick={0,0.0163810795111612,0.0327621590223223,0.0491432385334835,0.0655243180446447,0.0819053975558059,0.0913449294574825,0.0993259399615722,0.10623940572349,0.112337514702513,0.117792459066844,0.153679520577881,0.174672105822912,0.189566582088919,0.201119635432273,0.210559167333949,0.218540177838039,0.225453643599956,0.23155175257898,0.23700669694331,0.272893758454348},yticklabels={\(\displaystyle {0}\),\(\displaystyle {10^{-3}}\),,,,,,,,,\(\displaystyle {10^{-2}}\),,,,,,,,,\(\displaystyle {10^{-1}}\),},ylabel={Probability}},
colormap/viridis,
height=\figH,
point meta max=0.2925917804241,
point meta min=0,
tick align=outside,
tick pos=left,
width=\figW,
x grid style={white!69.0196078431373!black},
xlabel={Secret},
xmin=0, xmax=17,
xtick style={color=black},
y grid style={white!69.0196078431373!black},
ylabel={Time (cycles)},
ymin=464, ymax=538,
ytick style={color=black}
]
\addplot graphics [includegraphics cmd=\pgfimage,xmin=0, xmax=17, ymin=464, ymax=538] {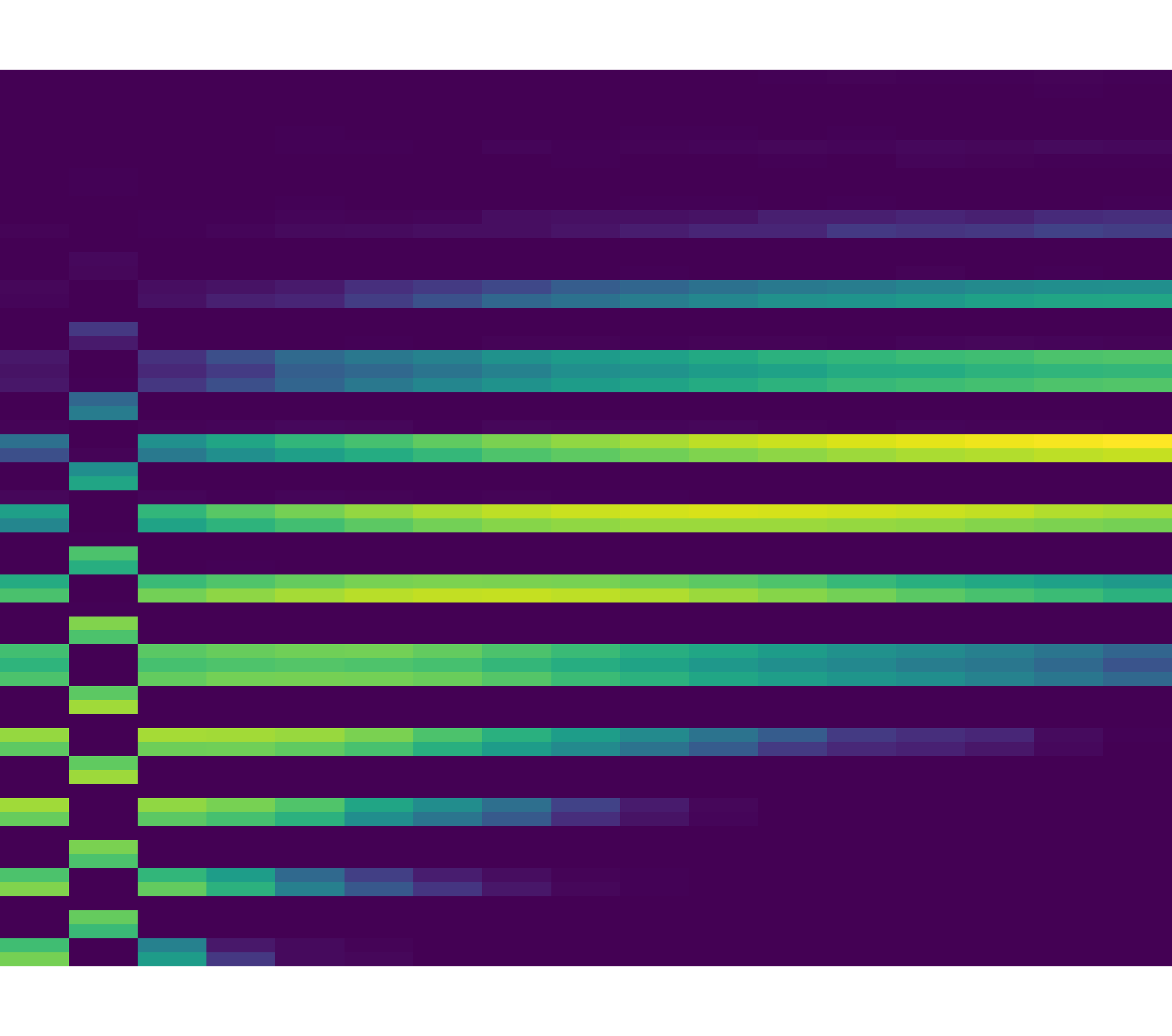};
\end{axis}

\end{tikzpicture}
        \caption{Unmitigated. \normalfont $N=10^6$, $\mathcal{M} = 408.7\text{mb}$, $\mathcal{M}_0 = 0.1\text{mb}$.}
    \end{subfigure}
    \hfill
    \begin{subfigure}[t]{0.45\linewidth}
\begin{tikzpicture}

\definecolor{color0}{rgb}{0.267004,0.004874,0.329415}

\begin{axis}[
axis background/.style={fill=color0},
colorbar,
colorbar style={ytick={0,0.052374576880886,0.075322633031377,0.0887463853438578,0.0982706891818681,0.105658313179764,0.111694441494349,0.116797915266621,0.121218745332359,0.12511819380683,0.128606369330255,0.151554425480746,0.164978177793227,0.174502481631237,0.181890105629133,0.187926233943718,0.19302970771599,0.197450537781728,0.201349986256199,0.204838161779624,0.227786217930115,0.241209970242596,0.250734274080606,0.258121898078502,0.264158026393087,0.269261500165359,0.273682330231097,0.277581778705567,0.281069954228993,0.304018010379484,0.317441762691965,0.326966066529975,0.334353690527871,0.340389818842456,0.345493292614728,0.349914122680466,0.353813571154936,0.357301746678362,0.380249802828853,0.393673555141334,0.403197858979344},yticklabels={\(\displaystyle {0}\),\(\displaystyle {10^{-5}}\),,,,,,,,,\(\displaystyle {10^{-4}}\),,,,,,,,,\(\displaystyle {10^{-3}}\),,,,,,,,,\(\displaystyle {10^{-2}}\),,,,,,,,,\(\displaystyle {10^{-1}}\),,,},ylabel={Probability}},
colormap/viridis,
height=\figH,
point meta max=0.4034850597382,
point meta min=0,
tick align=outside,
tick pos=left,
width=\figW,
x grid style={white!69.0196078431373!black},
xlabel={Secret},
xmin=0, xmax=17,
xtick style={color=black},
y grid style={white!69.0196078431373!black},
ylabel={Time (cycles)},
ymin=651, ymax=721,
ytick style={color=black}
]
\addplot graphics [includegraphics cmd=\pgfimage,xmin=0, xmax=17, ymin=651, ymax=721] {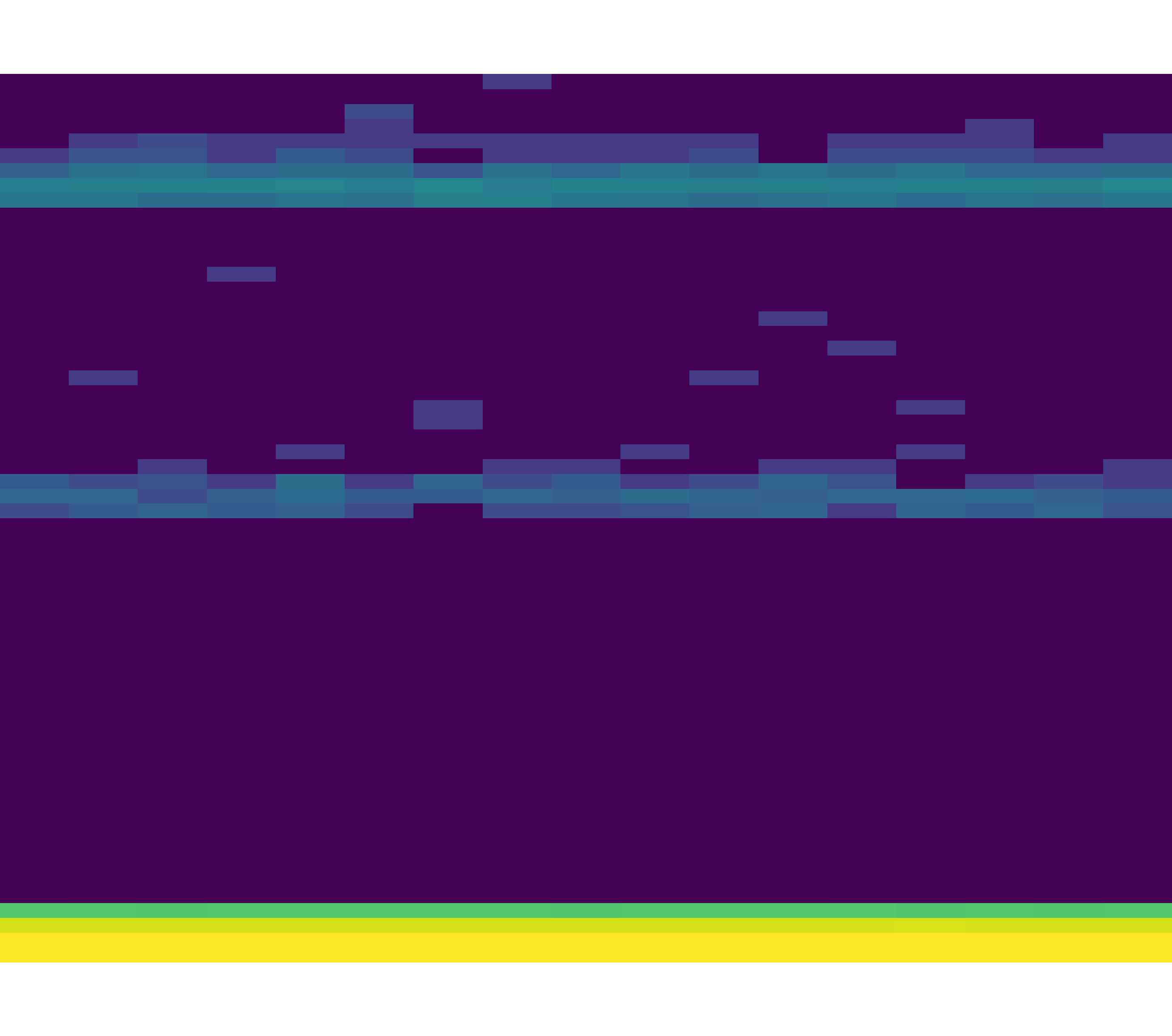};
\end{axis}

\end{tikzpicture}
        \caption{Improved fence (1). \normalfont $N=10^6$, $\mathcal{M} = 3.1\text{mb}$, $\mathcal{M}_0 = 7.7\text{mb}$.}
    \end{subfigure}
    \begin{subfigure}[t]{0.45\linewidth}
\begin{tikzpicture}

\definecolor{color0}{rgb}{0.267004,0.004874,0.329415}

\begin{axis}[
axis background/.style={fill=color0},
colorbar,
colorbar style={ytick={0,0.0981472239763183,0.141150683688187,0.166306095020904,0.184154143400057,0.197998165258664,0.209309554732773,0.21887319826394,0.227157603111926,0.234464966065489,0.241001624970533,0.284005084682402,0.309160496015119,0.327008544394271,0.340852566252879,0.352163955726988,0.361727599258155,0.370012004106141,0.377319367059704,0.383856025964748,0.426859485676617,0.452014897009334,0.469862945388486,0.483706967247094,0.495018356721203,0.50458200025237,0.512866405100355,0.520173768053919,0.526710426958963,0.569713886670832,0.594869298003548,0.612717346382701,0.626561368241309,0.637872757715418,0.647436401246585,0.65572080609457,0.663028169048134,0.669564827953178,0.712568287665047,0.737723698997763,0.755571747376916,0.769415769235523,0.780727158709632,0.790290802240799},yticklabels={\(\displaystyle {0}\),\(\displaystyle {10^{-5}}\),,,,,,,,,\(\displaystyle {10^{-4}}\),,,,,,,,,\(\displaystyle {10^{-3}}\),,,,,,,,,\(\displaystyle {10^{-2}}\),,,,,,,,,\(\displaystyle {10^{-1}}\),,,,,,},ylabel={Probability}},
colormap/viridis,
height=\figH,
point meta max=0.7984552979469,
point meta min=0,
tick align=outside,
tick pos=left,
width=\figW,
x grid style={white!69.0196078431373!black},
xlabel={Secret},
xmin=0, xmax=17,
xtick style={color=black},
y grid style={white!69.0196078431373!black},
ylabel={Time (cycles)},
ymin=651, ymax=725,
ytick style={color=black}
]
\addplot graphics [includegraphics cmd=\pgfimage,xmin=0, xmax=17, ymin=651, ymax=725] {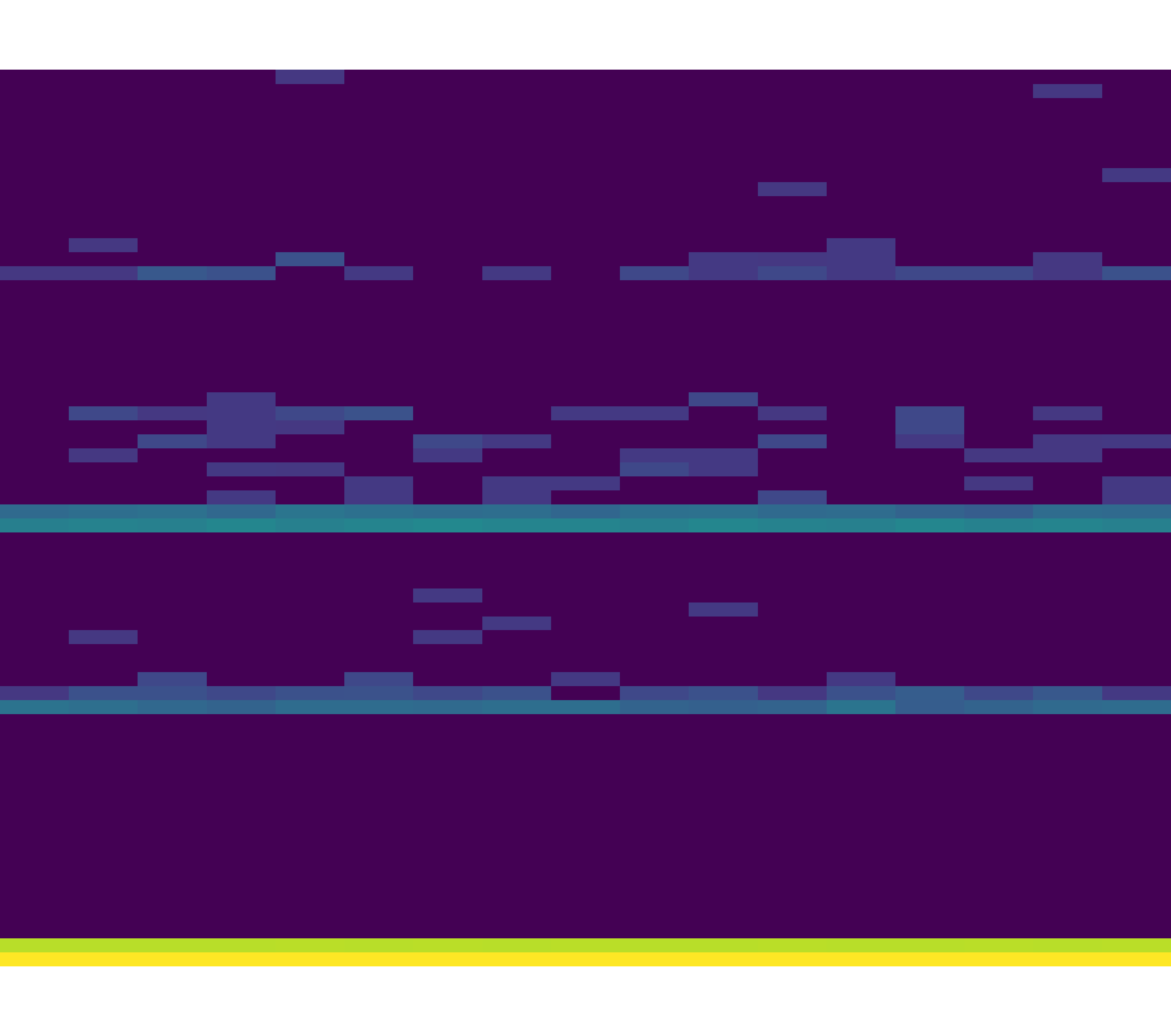};
\end{axis}

\end{tikzpicture}
        \caption{Improved fence (2). \normalfont $N=10^6$, $\mathcal{M} = 2.0\text{mb}$, $\mathcal{M}_0 = 7.2\text{mb}$.}
    \end{subfigure}
    \hfill
    \begin{subfigure}[t]{0.45\linewidth}
\begin{tikzpicture}

\definecolor{color0}{rgb}{0.267004,0.004874,0.329415}

\begin{axis}[
axis background/.style={fill=color0},
colorbar,
colorbar style={ytick={0,0.119472923279988,0.171820293228728,0.202441541660119,0.224167663177468,0.24101975225743,0.254788911608859,0.266430569962354,0.276515033126209,0.28541016004025,0.293367122206171,0.345714492154911,0.376335740586301,0.398061862103651,0.414913951183613,0.428683110535042,0.440324768888537,0.450409232052391,0.459304358966432,0.467261321132353,0.519608691081093,0.550229939512484,0.571956061029833,0.588808150109795,0.602577309461224,0.614218967814719,0.624303430978574,0.633198557892615,0.641155520058536,0.693502890007276,0.724124138438666,0.745850259956016,0.762702349035978,0.776471508387407,0.788113166740902,0.798197629904756,0.807092756818797,0.815049718984718,0.867397088933458,0.898018337364849,0.919744458882199,0.93659654796216,0.950365707313589,0.962007365667084,0.972091828830939,0.98098695574498},yticklabels={\(\displaystyle {0}\),\(\displaystyle {10^{-5}}\),,,,,,,,,\(\displaystyle {10^{-4}}\),,,,,,,,,\(\displaystyle {10^{-3}}\),,,,,,,,,\(\displaystyle {10^{-2}}\),,,,,,,,,\(\displaystyle {10^{-1}}\),,,,,,,,},ylabel={Probability}},
colormap/viridis,
height=\figH,
point meta max=0.9880348443985,
point meta min=0,
tick align=outside,
tick pos=left,
width=\figW,
x grid style={white!69.0196078431373!black},
xlabel={Secret},
xmin=0, xmax=17,
xtick style={color=black},
y grid style={white!69.0196078431373!black},
ylabel={Time (cycles)},
ymin=651, ymax=788,
ytick style={color=black}
]
\addplot graphics [includegraphics cmd=\pgfimage,xmin=0, xmax=17, ymin=651, ymax=788] {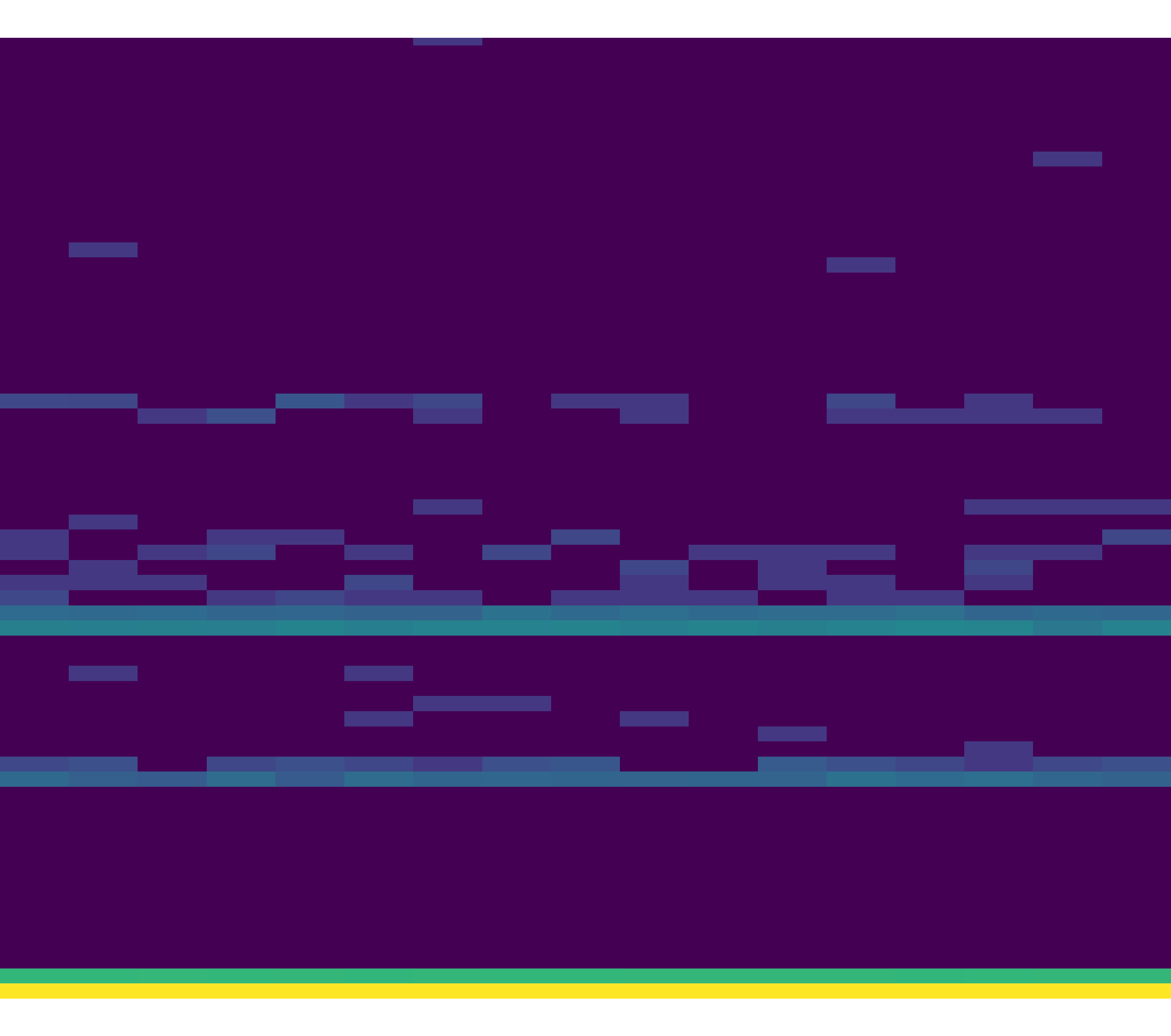};
\end{axis}

\end{tikzpicture}
        \caption{Improved fence (3). \normalfont $N=10^6$, $\mathcal{M} = 1.7\text{mb}$, $\mathcal{M}_0 = 8.6\text{mb}$.}
    \end{subfigure}
    \caption{TLB channel matrices.}
    \label{fig:tlb}
\end{figure*}

\begin{figure*}
    \captionsetup[subfigure]{aboveskip=-2ex,belowskip=1ex}
    \begin{subfigure}[t]{0.45\linewidth}
\begin{tikzpicture}

\definecolor{color0}{rgb}{0.267004,0.004874,0.329415}

\begin{axis}[
axis background/.style={fill=color0},
colorbar,
colorbar style={ytick={0,0.146235649485925,0.21030917705329,0.247789787969122,0.274382704620656,0.29500977328313,0.311863315536488,0.326112782475762,0.338456232188021,0.34934392645232,0.359083300850496,0.423156828417861,0.460637439333693,0.487230355985226,0.507857424647701,0.524710966901058,0.538960433840332,0.551303883552592,0.56219157781689,0.571930952215066,0.636004479782432,0.673485090698263,0.700078007349797,0.720705076012272,0.737558618265629,0.751808085204902,0.764151534917162,0.775039229181461,0.784778603579637,0.848852131147002,0.886332742062834,0.912925658714367,0.933552727376842,0.950406269630199,0.964655736569473,0.976999186281733,0.987886880546031},yticklabels={\(\displaystyle {0}\),\(\displaystyle {10^{-4}}\),,,,,,,,,\(\displaystyle {10^{-3}}\),,,,,,,,,\(\displaystyle {10^{-2}}\),,,,,,,,,\(\displaystyle {10^{-1}}\),,,,,,,,},ylabel={Probability}},
colormap/viridis,
height=\figH,
point meta max=0.9973841309547,
point meta min=0,
tick align=outside,
tick pos=left,
width=\figW,
x grid style={white!69.0196078431373!black},
xlabel={Secret},
xmin=0, xmax=17,
xtick style={color=black},
y grid style={white!69.0196078431373!black},
ylabel={Time (cycles)},
ymin=69, ymax=206,
ytick style={color=black}
]
\addplot graphics [includegraphics cmd=\pgfimage,xmin=0, xmax=17, ymin=69, ymax=206] {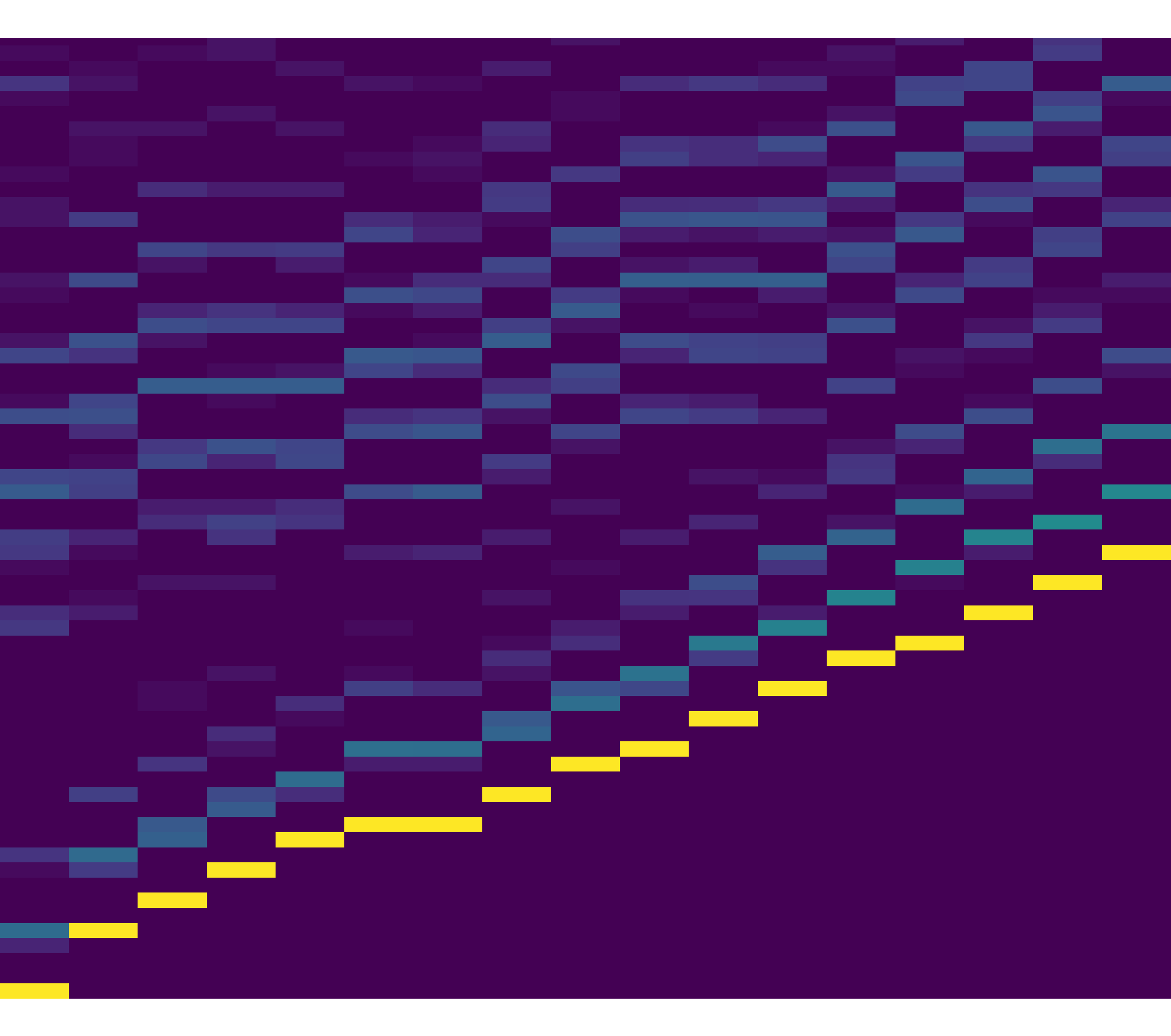};
\end{axis}

\end{tikzpicture}
        \caption{Unmitigated. \normalfont $N=10^6$, $\mathcal{M} = 3211.4\text{mb}$, $\mathcal{M}_0 = 0.1\text{mb}$.}
    \end{subfigure}
    \hfill
    \begin{subfigure}[t]{0.45\linewidth}
\begin{tikzpicture}

\definecolor{color0}{rgb}{0.267004,0.004874,0.329415}

\begin{axis}[
axis background/.style={fill=color0},
colorbar,
colorbar style={ytick={0,0.120778125104349,0.173697372603933,0.204653147957572,0.226616620103518,0.243652812633932,0.257572395457157,0.269341235043892,0.279535867603103,0.288528170810796,0.296572060133517,0.349491307633101,0.38044708298674,0.402410555132686,0.4194467476631,0.433366330486325,0.44513517007306,0.455329802632271,0.464322105839964,0.472365995162685,0.525285242662269,0.556241018015908,0.578204490161854,0.595240682692268,0.609160265515493,0.620929105102228,0.631123737661439,0.640116040869132,0.648159930191853,0.701079177691437,0.732034953045076,0.753998425191022,0.771034617721436,0.784954200544661,0.796723040131396,0.806917672690607,0.8159099758983,0.823953865221021,0.876873112720605,0.907828888074244,0.92979236022019,0.946828552750604,0.960748135573829,0.972516975160564,0.982711607719775,0.991703910927468},yticklabels={\(\displaystyle {0}\),\(\displaystyle {10^{-5}}\),,,,,,,,,\(\displaystyle {10^{-4}}\),,,,,,,,,\(\displaystyle {10^{-3}}\),,,,,,,,,\(\displaystyle {10^{-2}}\),,,,,,,,,\(\displaystyle {10^{-1}}\),,,,,,,,},ylabel={Probability}},
colormap/viridis,
height=\figH,
point meta max=0.9997269511223,
point meta min=0,
tick align=outside,
tick pos=left,
width=\figW,
x grid style={white!69.0196078431373!black},
xlabel={Secret},
xmin=0, xmax=17,
xtick style={color=black},
y grid style={white!69.0196078431373!black},
ylabel={Time (cycles)},
ymin=426, ymax=491,
ytick style={color=black}
]
\addplot graphics [includegraphics cmd=\pgfimage,xmin=0, xmax=17, ymin=426, ymax=491] {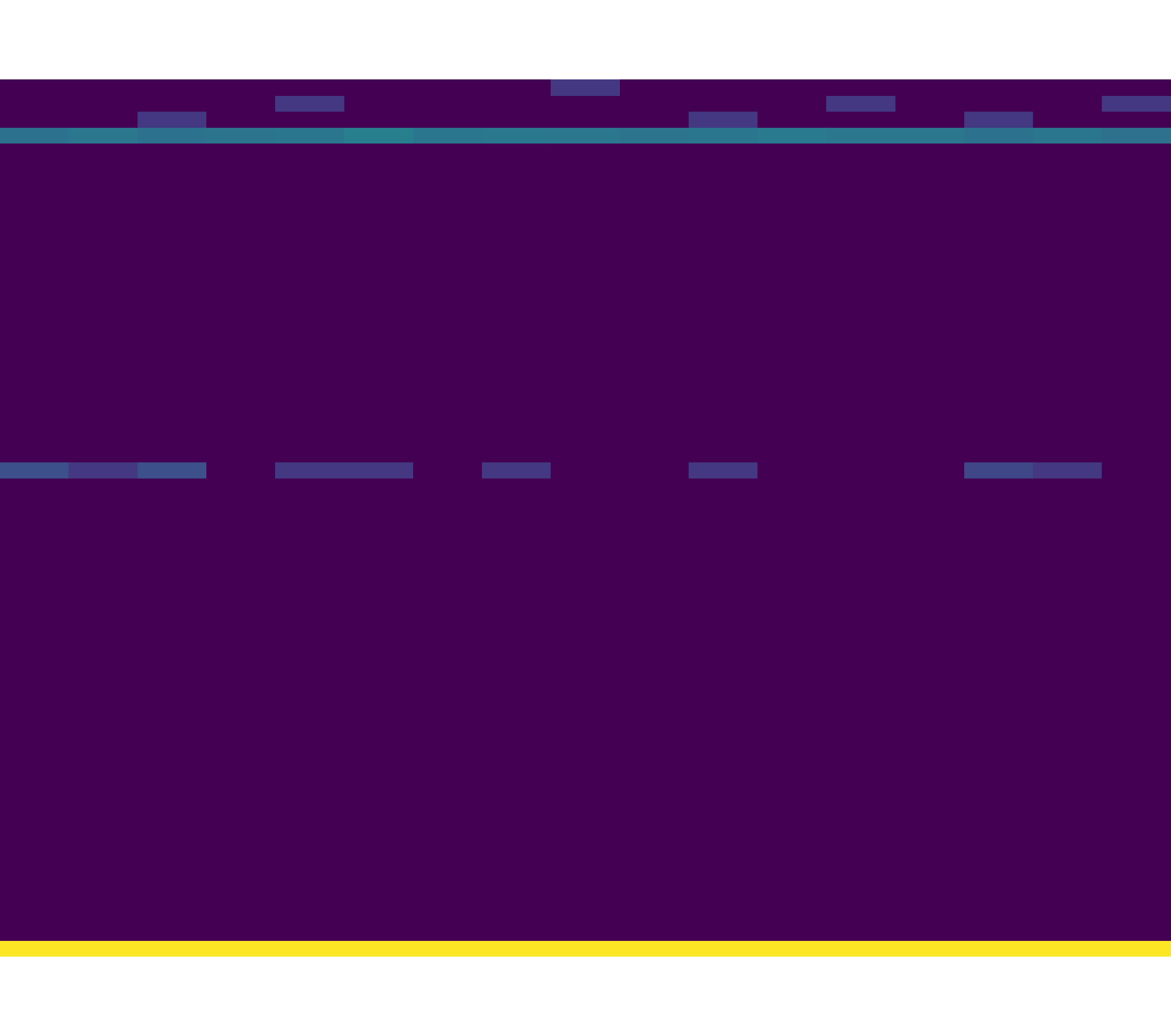};
\end{axis}

\end{tikzpicture}
        \caption{Improved fence (1). \normalfont $N=10^6$, $\mathcal{M} = 28.2\text{mb}$, $\mathcal{M}_0 = 60.3\text{mb}$.}
    \end{subfigure}
    \begin{subfigure}[t]{0.45\linewidth}
\begin{tikzpicture}

\definecolor{color0}{rgb}{0.267004,0.004874,0.329415}

\begin{axis}[
axis background/.style={fill=color0},
colorbar,
colorbar style={ytick={0,0.120778351238014,0.173697697818648,0.204653531130985,0.226617044399282,0.24365326882667,0.257572877711619,0.269341739333229,0.279536390979916,0.288528711023957,0.296572615407304,0.349491961987938,0.380447795300275,0.402411308568572,0.41944753299596,0.433367141880909,0.445136003502519,0.455330655149206,0.464322975193247,0.472366879576594,0.525286226157228,0.556242059469565,0.578205572737862,0.59524179716525,0.609161406050199,0.620930267671809,0.631124919318496,0.640117239362537,0.648161143745884,0.701080490326518,0.732036323638855,0.753999836907152,0.77103606133454,0.784955670219489,0.796724531841099,0.806919183487786,0.815911503531827,0.823955407915174,0.876874754495808,0.907830587808145,0.929794101076442,0.94683032550383,0.960749934388779,0.972518796010389,0.982713447657076,0.991705767701117},yticklabels={\(\displaystyle {0}\),\(\displaystyle {10^{-5}}\),,,,,,,,,\(\displaystyle {10^{-4}}\),,,,,,,,,\(\displaystyle {10^{-3}}\),,,,,,,,,\(\displaystyle {10^{-2}}\),,,,,,,,,\(\displaystyle {10^{-1}}\),,,,,,,,},ylabel={Probability}},
colormap/viridis,
height=\figH,
point meta max=0.9997289776802,
point meta min=0,
tick align=outside,
tick pos=left,
width=\figW,
x grid style={white!69.0196078431373!black},
xlabel={Secret},
xmin=0, xmax=17,
xtick style={color=black},
y grid style={white!69.0196078431373!black},
ylabel={Time (cycles)},
ymin=426, ymax=500,
ytick style={color=black}
]
\addplot graphics [includegraphics cmd=\pgfimage,xmin=0, xmax=17, ymin=426, ymax=500] {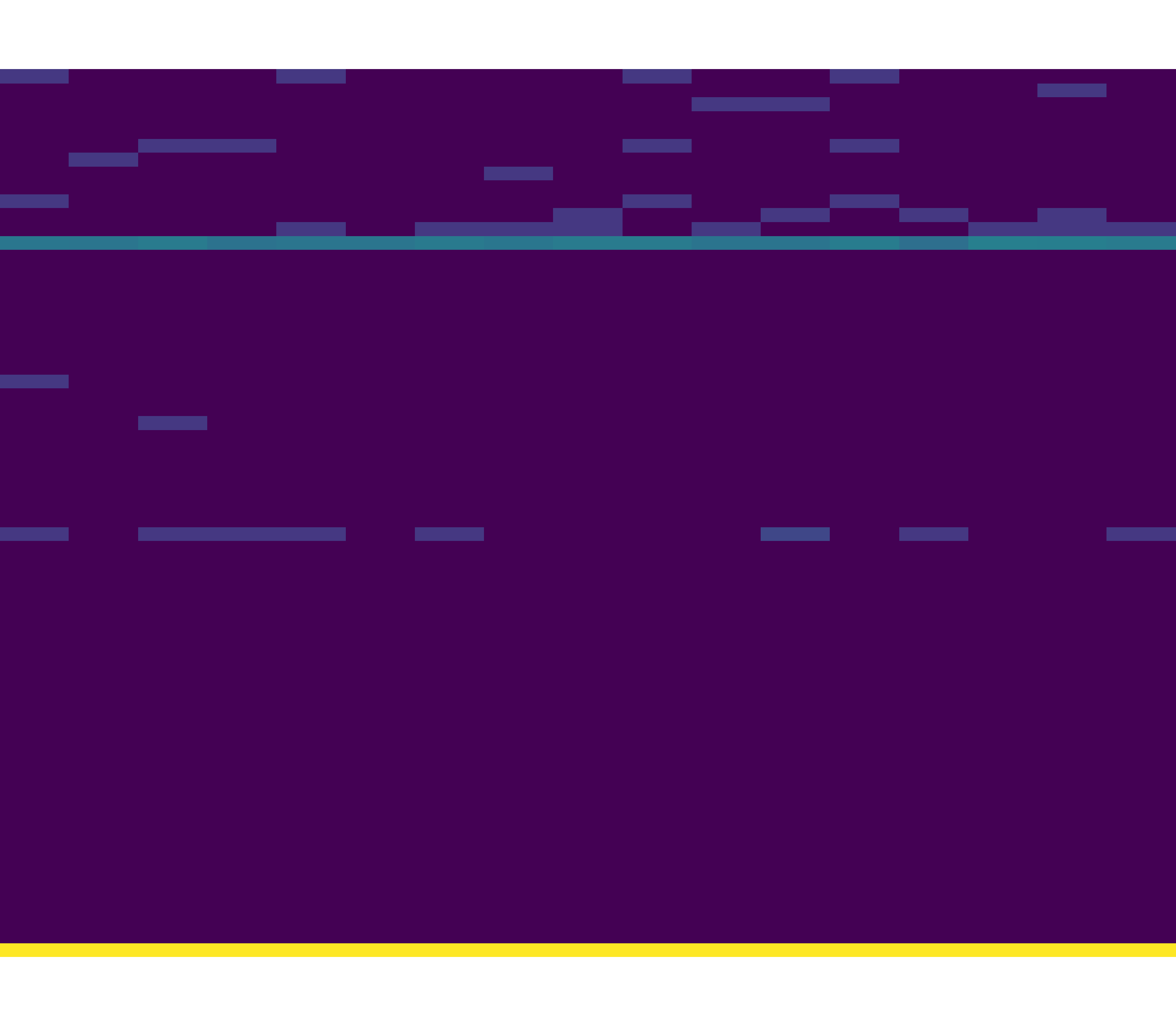};
\end{axis}

\end{tikzpicture}
        \caption{Improved fence (2). \normalfont $N=10^6$, $\mathcal{M} = 43.8\text{mb}$, $\mathcal{M}_0 = 58.0\text{mb}$.}
    \end{subfigure}
    \hfill
    \begin{subfigure}[t]{0.45\linewidth}
\begin{tikzpicture}

\definecolor{color0}{rgb}{0.267004,0.004874,0.329415}

\begin{axis}[
axis background/.style={fill=color0},
colorbar,
colorbar style={ytick={0,0.120772624725547,0.173689462219952,0.204643827810934,0.226606299714356,0.243641716396394,0.257560665305338,0.269328968925186,0.279523137208761,0.28851503089632,0.296558553890799,0.349475391385203,0.380429756976185,0.402392228879607,0.419427645561645,0.433346594470589,0.445114898090437,0.455309066374012,0.464300960061571,0.47234448305605,0.525261320550454,0.556215686141436,0.578178158044859,0.595213574726896,0.609132523635841,0.620900827255689,0.631094995539263,0.640086889226823,0.648130412221301,0.701047249715706,0.732001615306688,0.75396408721011,0.770999503892148,0.784918452801092,0.79668675642094,0.806880924704515,0.815872818392074,0.823916341386552,0.876833178880957,0.907787544471939,0.929750016375361,0.946785433057399,0.960704381966343,0.972472685586191,0.982666853869766,0.991658747557325},yticklabels={\(\displaystyle {0}\),\(\displaystyle {10^{-5}}\),,,,,,,,,\(\displaystyle {10^{-4}}\),,,,,,,,,\(\displaystyle {10^{-3}}\),,,,,,,,,\(\displaystyle {10^{-2}}\),,,,,,,,,\(\displaystyle {10^{-1}}\),,,,,,,,},ylabel={Probability}},
colormap/viridis,
height=\figH,
point meta max=0.9996776580811,
point meta min=0,
tick align=outside,
tick pos=left,
width=\figW,
x grid style={white!69.0196078431373!black},
xlabel={Secret},
xmin=0, xmax=17,
xtick style={color=black},
y grid style={white!69.0196078431373!black},
ylabel={Time (cycles)},
ymin=426, ymax=500,
ytick style={color=black}
]
\addplot graphics [includegraphics cmd=\pgfimage,xmin=0, xmax=17, ymin=426, ymax=500] {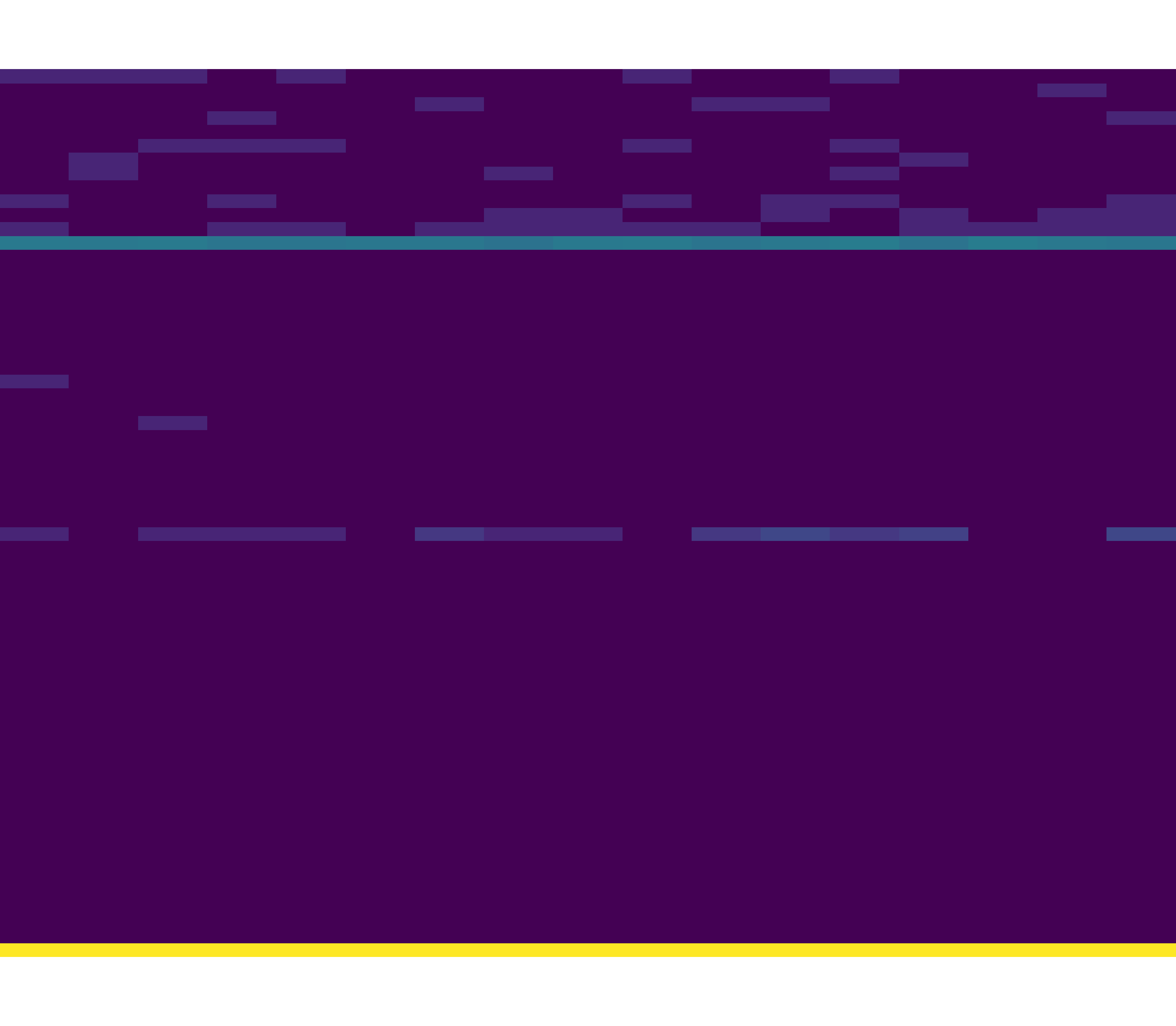};
\end{axis}

\end{tikzpicture}
        \caption{Improved fence (3). \normalfont $N=10^6$, $\mathcal{M} = 20.1\text{mb}$, $\mathcal{M}_0 = 35.3\text{mb}$.}
    \end{subfigure}
    \caption{BTB channel matrices.}
    \label{fig:btb}
\end{figure*}

\pgfplotsset{every axis/.append style={xtick distance=16}}
\begin{figure*}[h]
    \captionsetup[subfigure]{aboveskip=-2ex,belowskip=1ex}
    \begin{subfigure}[t]{0.45\linewidth}
\begin{tikzpicture}

\definecolor{color0}{rgb}{0.267004,0.004874,0.329415}

\begin{axis}[
axis background/.style={fill=color0},
colorbar,
colorbar style={ytick={0,0.146253260204356,0.210334503953492,0.247819628546308,0.274415747702628,0.2950453004208,0.311900872295445,0.326152055255066,0.338496991451764,0.349385996888261,0.359126544169936,0.423207787919073,0.460692912511889,0.487289031668209,0.507918584386381,0.524774156261025,0.539025339220647,0.551370275417345,0.562259280853842,0.571999828135517,0.636081071884653,0.67356619647747,0.70016231563379,0.720791868351962,0.737647440226606,0.751898623186228,0.764243559382926,0.775132564819423,0.784873112101098,0.848954355850234,0.886439480443051,0.91303559959937,0.933665152317543,0.950520724192187,0.964771907151808,0.977116843348507,0.988005848785004},yticklabels={\(\displaystyle {0}\),\(\displaystyle {10^{-4}}\),,,,,,,,,\(\displaystyle {10^{-3}}\),,,,,,,,,\(\displaystyle {10^{-2}}\),,,,,,,,,\(\displaystyle {10^{-1}}\),,,,,,,,},ylabel={Probability}},
colormap/viridis,
height=\figH,
point meta max=0.9975165128708,
point meta min=0,
tick align=outside,
tick pos=left,
width=\figW,
x grid style={white!69.0196078431373!black},
xlabel={Secret},
xmin=0, xmax=65,
xtick style={color=black},
y grid style={white!69.0196078431373!black},
ylabel={Time (cycles)},
ymin=162, ymax=488,
ytick style={color=black}
]
\addplot graphics [includegraphics cmd=\pgfimage,xmin=0, xmax=65, ymin=162, ymax=488] {figures/bht_sm-001.png};
\end{axis}

\end{tikzpicture}
        \caption{Unmitigated. \normalfont $N=10^6$, $\mathcal{M} = 3770.6\text{mb}$, $\mathcal{M}_0 = 0.2\text{mb}$.}
    \end{subfigure}
    \hfill
    \begin{subfigure}[t]{0.45\linewidth}
\begin{tikzpicture}

\definecolor{color0}{rgb}{0.267004,0.004874,0.329415}

\begin{axis}[
axis background/.style={fill=color0},
colorbar,
colorbar style={ytick={0,0.120721976450847,0.173616622272861,0.204558006567666,0.226511268094875,0.243539540654098,0.25745265238968,0.269216020749772,0.279405913916889,0.288394036684485,0.296434186476112,0.349328832298125,0.380270216592931,0.402223478120139,0.419251750679362,0.433164862414944,0.444928230775036,0.455118123942153,0.46410624670975,0.472146396501376,0.52504104232339,0.555982426618195,0.577935688145404,0.594963960704626,0.608877072440209,0.6206404408003,0.630830333967417,0.639818456735014,0.64785860652664,0.700753252348654,0.731694636643459,0.753647898170668,0.770676170729891,0.784589282465473,0.796352650825565,0.806542543992682,0.815530666760278,0.823570816551905,0.876465462373918,0.907406846668724,0.929360108195932,0.946388380755155,0.960301492490737,0.972064860850829,0.982254754017946,0.991242876785542},yticklabels={\(\displaystyle {0}\),\(\displaystyle {10^{-5}}\),,,,,,,,,\(\displaystyle {10^{-4}}\),,,,,,,,,\(\displaystyle {10^{-3}}\),,,,,,,,,\(\displaystyle {10^{-2}}\),,,,,,,,,\(\displaystyle {10^{-1}}\),,,,,,,,},ylabel={Probability}},
colormap/viridis,
height=\figH,
point meta max=0.9992237687111,
point meta min=0,
tick align=outside,
tick pos=left,
width=\figW,
x grid style={white!69.0196078431373!black},
xlabel={Secret},
xmin=0, xmax=65,
xtick style={color=black},
y grid style={white!69.0196078431373!black},
ylabel={Time (cycles)},
ymin=1197, ymax=1271,
ytick style={color=black}
]
\addplot graphics [includegraphics cmd=\pgfimage,xmin=0, xmax=65, ymin=1197, ymax=1271] {figures/bht_fencet_sm-000.png};
\end{axis}

\end{tikzpicture}
        \caption{Improved fence (1). \normalfont $N=10^6$, $\mathcal{M} = 44.1\text{mb}$, $\mathcal{M}_0 = 60.8\text{mb}$.}
    \end{subfigure}
    \begin{subfigure}[t]{0.45\linewidth}
\begin{tikzpicture}

\definecolor{color0}{rgb}{0.267004,0.004874,0.329415}

\begin{axis}[
axis background/.style={fill=color0},
colorbar,
colorbar style={ytick={0,0.12074345311236,0.173647508989093,0.204594397813038,0.226551564865826,0.243582866786037,0.257498453689771,0.269263914774716,0.279455620742558,0.288445342513716,0.29648692266277,0.349390978539502,0.380337867363448,0.402295034416235,0.419326336336446,0.43324192324018,0.445007384325126,0.455199090292968,0.464188812064126,0.472230392213179,0.525134448089912,0.556081336913857,0.578038503966644,0.595069805886855,0.60898539279059,0.620750853875535,0.630942559843377,0.639932281614535,0.647973861763588,0.700877917640321,0.731824806464266,0.753781973517054,0.770813275437265,0.784728862340999,0.796494323425944,0.806686029393787,0.815675751164945,0.823717331313998,0.87662138719073,0.907568276014676,0.929525443067463,0.946556744987674,0.960472331891408,0.972237792976354,0.982429498944196,0.991419220715353},yticklabels={\(\displaystyle {0}\),\(\displaystyle {10^{-5}}\),,,,,,,,,\(\displaystyle {10^{-4}}\),,,,,,,,,\(\displaystyle {10^{-3}}\),,,,,,,,,\(\displaystyle {10^{-2}}\),,,,,,,,,\(\displaystyle {10^{-1}}\),,,,,,,,},ylabel={Probability}},
colormap/viridis,
height=\figH,
point meta max=0.9994162321091,
point meta min=0,
tick align=outside,
tick pos=left,
width=\figW,
x grid style={white!69.0196078431373!black},
xlabel={Secret},
xmin=0, xmax=65,
xtick style={color=black},
y grid style={white!69.0196078431373!black},
ylabel={Time (cycles)},
ymin=1197, ymax=1271,
ytick style={color=black}
]
\addplot graphics [includegraphics cmd=\pgfimage,xmin=0, xmax=65, ymin=1197, ymax=1271] {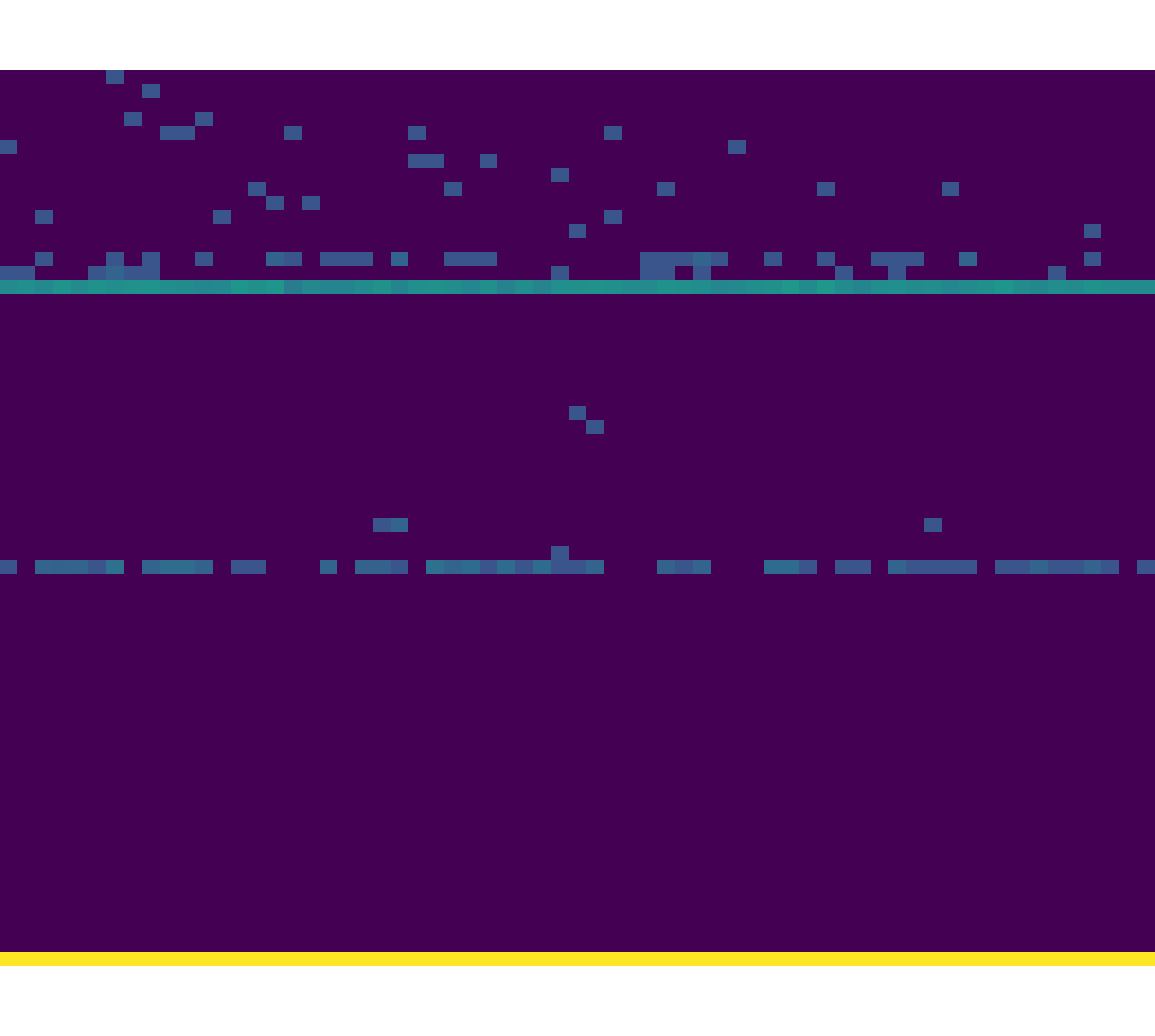};
\end{axis}

\end{tikzpicture}
        \caption{Improved fence (2). \normalfont $N=10^6$, $\mathcal{M} = 49.8\text{mb}$, $\mathcal{M}_0 = 56.1\text{mb}$.}
    \end{subfigure}
    \hfill
    \begin{subfigure}[t]{0.45\linewidth}
\begin{tikzpicture}

\definecolor{color0}{rgb}{0.267004,0.004874,0.329415}

\begin{axis}[
axis background/.style={fill=color0},
colorbar,
colorbar style={ytick={0,0.120722096172897,0.173616794451436,0.204558209431342,0.226511492729975,0.243539782176427,0.257452907709881,0.269216287735906,0.279406191008514,0.288394322689786,0.296434480454967,0.349329178733506,0.380270593713411,0.402223877012045,0.419252166458497,0.43316529199195,0.444928672017975,0.455118575290584,0.464106706971856,0.472146864737036,0.525041563015575,0.555982977995481,0.577936261294114,0.594964550740566,0.60887767627402,0.620641056300045,0.630830959572654,0.639819091253925,0.647859249019106,0.700753947297645,0.73169536227755,0.753648645576184,0.770676935022636,0.784590060556089,0.796353440582115,0.806543343854723,0.815531475535995,0.823571633301176,0.876466331579714,0.90740774655962,0.929361029858254,0.946389319304706,0.960302444838159,0.972065824864184,0.982255728136793,0.991243859818064},yticklabels={\(\displaystyle {0}\),\(\displaystyle {10^{-5}}\),,,,,,,,,\(\displaystyle {10^{-4}}\),,,,,,,,,\(\displaystyle {10^{-3}}\),,,,,,,,,\(\displaystyle {10^{-2}}\),,,,,,,,,\(\displaystyle {10^{-1}}\),,,,,,,,},ylabel={Probability}},
colormap/viridis,
height=\figH,
point meta max=0.9992248415947,
point meta min=0,
tick align=outside,
tick pos=left,
width=\figW,
x grid style={white!69.0196078431373!black},
xlabel={Secret},
xmin=0, xmax=65,
xtick style={color=black},
y grid style={white!69.0196078431373!black},
ylabel={Time (cycles)},
ymin=1197, ymax=1271,
ytick style={color=black}
]
\addplot graphics [includegraphics cmd=\pgfimage,xmin=0, xmax=65, ymin=1197, ymax=1271] {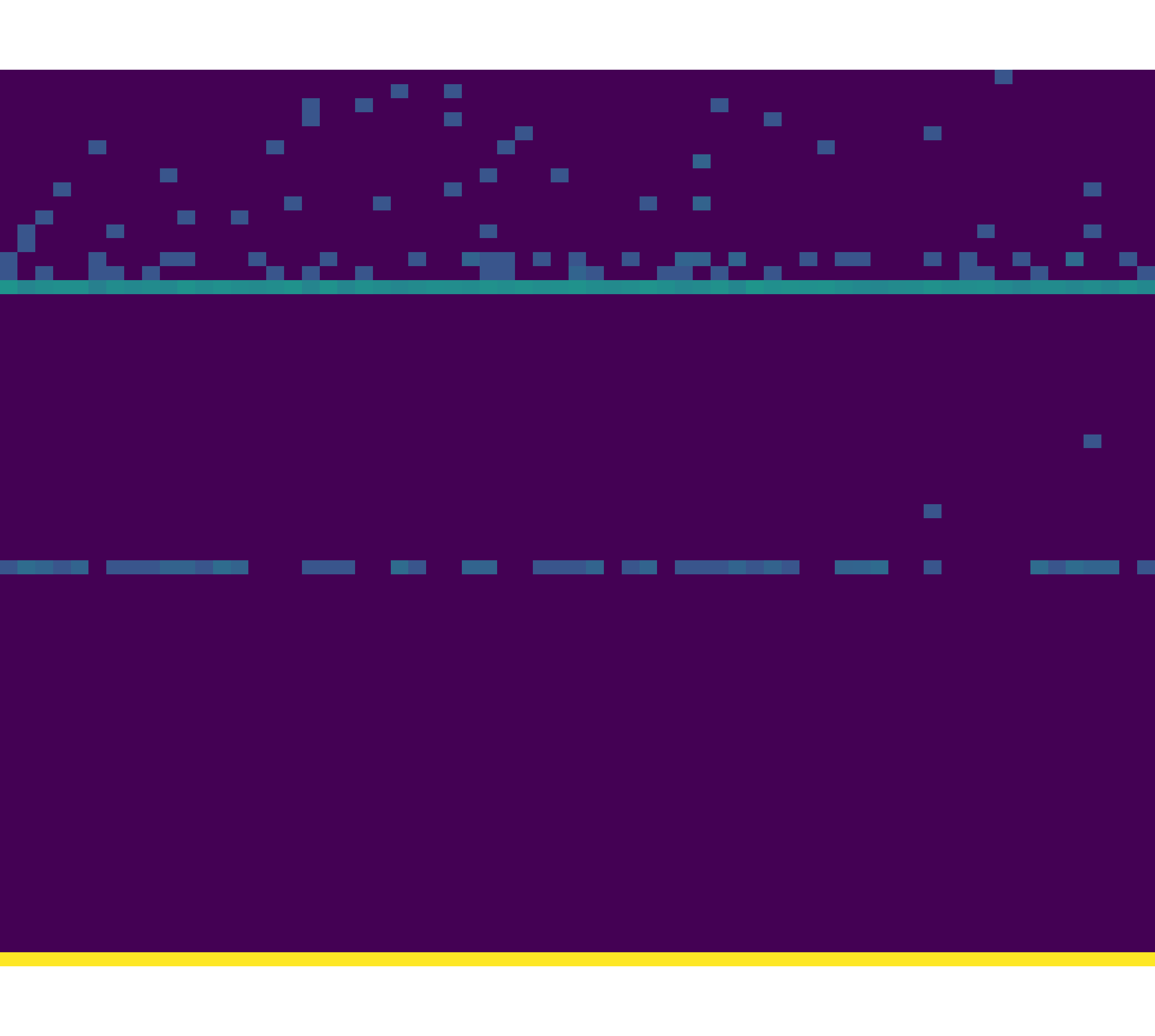};
\end{axis}

\end{tikzpicture}
        \caption{Improved fence (3). \normalfont $N=10^6$, $\mathcal{M} = 44.6\text{mb}$, $\mathcal{M}_0 = 58.6\text{mb}$.}
    \end{subfigure}
    \caption{BHT channel matrices.}
    \label{fig:bht-full}
\end{figure*}

\twocolumn
\fi 

\end{document}